\documentclass[12pt,onecolumn]{IEEEtran}

\usepackage[intlimits]{amsmath} 
\usepackage{amsxtra}     
\usepackage{amsthm}
\usepackage{amssymb}
\usepackage{amsfonts}
\usepackage{amsmath}
\usepackage{dsfont}
\usepackage{graphicx}    
\usepackage{multirow}
\usepackage{epsfig}
\usepackage{subfigure}   
\usepackage{verbatim}
\usepackage{dbl12}
\usepackage{cite}
\usepackage{bm}
\usepackage[usenames]{color}
\usepackage{algorithm}
\usepackage{algpseudocode}
\usepackage{colortbl}
\usepackage{rotating}
\usepackage{url}
\usepackage{dblfloatfix}

\newcommand{\detectionperffigswitch}[2]{#1}
\newcommand{\ignoreForJournal}[1]{#1}
\newcommand{\journalTechSwitch}[2]{#2}
\newcommand{\perfpredswitch}[2]{}
\newcommand{\forMastersThesis}[1]{}

\newcommand{\norm}[1]{\left\|#1\right\|}
\newcommand{\set}[1]{\left\{#1\right\}}
\newcommand{\ignore}[1]{}

\newcommand{\vq}{{\bm q}}
\newcommand{\vdotq}{\dot{\vq}}
\newcommand{\vr}{{\bm r}}
\newcommand{\vp}{{\bm p}}
\newcommand{\vdotr}{\dot{\vr}}

\newcommand{\calX}{{\cal X}}

\newcommand{\sarmat}[1]{{\bm{#1}}_{f,i}}
\newcommand{\sarImat}{\sarmat{I}}

\newcommand{\sarHmat}{\sarmat{H}}
\newcommand{\sarLmat}{\sarmat{L}}
\newcommand{\sarSmat}{\sarmat{S}}
\newcommand{\sarBmat}{{\bm B}_{f}}
\newcommand{\sarXmat}{\sarmat{X}}
\newcommand{\sarGmat}{\sarmat{G}}
\newcommand{\sarMmat}{\sarmat{M}}
\newcommand{\sarVmat}{\sarmat{V}}

\newcommand{\sarDmatG}{{\bm \Delta}_f^G}
\newcommand{\sarDmatM}{{\bm \Delta}_{f,i}^M}

\newcommand{\sarXvi}{{\bm x}_{f,1:N}^{(p)}}
\newcommand{\sarGvi}{{\bm g}_{f,1:N}^{(p)}}
\newcommand{\sarDMi}{{\bm \delta}_{f,1:N}^{M,(p)}}
\newcommand{\sarDG}{\delta_{f}^{G,(p)}}
\newcommand{\sarDM}{\delta_{f,i}^{M,(p)}}
\newcommand{\sarMvi}{{\bm m}_{f,1:N}^{(p)}}

\ignore{\newcommand{\sarHv}{\vec{h}_{f,i}^{(p)}}
\newcommand{\sarIv}{\vec{i}_{f,i}^{(p)}}
\newcommand{\sarRv}{\vec{r}_{f,i}^{(p)}}
\newcommand{\sarRvf}{\vec{r}_{f}^{(p)}}
\newcommand{\sarVv}{\vec{v}_{f,i}^{(p)}}
\newcommand{\sarBv}{\vec{b}_{f}^{(p)}}
\newcommand{\sarMv}{\vec{m}_{f,i}^{(p)}}
\newcommand{\sarXv}{\vec{x}_{f,i}^{(p)}}
\newcommand{\sarGv}{\vec{g}_{f,i}^{(p)}}}

\newcommand{\sarHv}{\underline{h}_{f,i}^{(p)}}
\newcommand{\sarIv}{\underline{i}_{f,i}^{(p)}}
\newcommand{\sarRv}{\underline{r}_{f,i}^{(p)}}
\newcommand{\sarRvf}{\underline{r}_{f}^{(p)}}
\newcommand{\sarVv}{\underline{v}_{f,i}^{(p)}}
\newcommand{\sarBv}{\underline{b}_{f}^{(p)}}
\newcommand{\sarMv}{\underline{m}_{f,i}^{(p)}}
\newcommand{\sarXv}{\underline{x}_{f,i}^{(p)}}
\newcommand{\sarGv}{\underline{g}_{f,i}^{(p)}}

\newcommand{\sarH}{H_{k,f,i}(p)}

\newcommand{\sarV}{V_{k,f,i}(p)}

\newcommand{\GammaB}{{\bm \Gamma}^B}
\newcommand{\GammaX}{{\bm \Gamma}^X}
\newcommand{\GammaG}{{\bm \Gamma}^G}
\newcommand{\GammaM}{{\bm \Gamma}^M}
\newcommand{\GammaV}{{\bm \Gamma}^V}

\newcommand{\cn}[2]{\mathcal{CN}\left(#1,#2\right)}
\newcommand{\normpdf}{\phi_{\mathcal{CN}}}
\ignore{
\definecolor{defaultlinecolor}{named}{Black}
\definecolor{linecolor}{named}{White}
\definecolor{textRowHeader}{named}{White}
\definecolor{textColumnHeader}{named}{White}
\newcolumntype{A}{>{\color{textRowHeader}\columncolor{firstrow}}c}	
\newcolumntype{B}{>{\color{textColumnHeader}\columncolor{firstcol}}l}	
\newcolumntype{C}{>{\color{textColumnHeader}\columncolor{firstsubcol}}l}	
\newcolumntype{D}{>{\color{textColumnHeader}\columncolor{firstcol}}c}	
\newcolumntype{E}{>{\color{textColumnHeader}\columncolor{firstsubcol}}c}	

\newcommand{\firstrowfirstcell}[1]{\multicolumn{1}{A}{\bf #1}}
\newcommand{\firstrowcell}[1]{\multicolumn{1}{A}{\bf #1}}

\newcommand{\firstcolcellc}[1]{\multicolumn{1}{D}{#1}}

\newcommand{\oddTableRowc}[2]{\rowcolor[gray]{0.9}\firstcolcellc{#1} & #2}

\newcommand{\evenTableRowc}[2]{\rowcolor[gray]{0.75}\firstcolcellc{#1} & #2}

}

\newenvironment{changemargin}[2]{%
  \begin{list}{}{%
    \setlength{\topsep}{0pt}%
    \setlength{\leftmargin}{#1}%
    \setlength{\rightmargin}{#2}%
    \setlength{\listparindent}{\parindent}%
    \setlength{\itemindent}{\parindent}%
    \setlength{\parsep}{\parskip}%
  }%
  \item[]}{\end{list}}

\newcommand{\trieq}{\stackrel{\bigtriangleup}{=}}

\newcommand{\intinf}[0]{\int_{-\infty}^{\infty}}

\newcommand{\oneK}{{\bm 1}_K}
\newcommand{\eyeKK}{{\bm I}_{K\times K}}

\theoremstyle{plain}

\theoremstyle{definition}

\begin{document}
\title{Moving target inference with hierarchical Bayesian models in synthetic aperture radar imagery}

\author{Gregory~E.~Newstadt,~\IEEEmembership{Student~Member,~IEEE,} ~Edmund~G.~Zelnio, and~Alfred~O.~Hero III,~\IEEEmembership{Fellow,~IEEE}
\thanks{Gregory Newstadt and Alfred Hero are with the Dept. of Electrical Engineering and Computer Science, University of Michigan, Ann
Arbor.  Edmund Zelnio is with the Air Force Research Laboratory, Wright Patterson Air Force Base, OH 45433, USA. E-mail: (\{newstage\},\{hero\}@umich.edu and edmund.zelnio@wpafb.af.mil).}
\thanks{The research in this paper was partially supported by Air Force Office of Scientific Research award FA9550-06-1-0324 and by Air Force Research Laboratory award FA8650-07-D-1221-TO1.}
\thanks{This document was cleared for public release under document number 88ABW-2013-0611.}
\\ \today}

\maketitle

\begin{abstract}
In synthetic aperture radar (SAR), images are formed by focusing the response of stationary objects to a single spatial location. On the other hand, moving targets cause phase errors in the standard formation of SAR images that cause displacement and defocusing effects.  SAR imagery also contains significant sources of non-stationary spatially-varying noises, including antenna gain discrepancies, angular scintillation (glints) and complex speckle.  In order to account for this intricate phenomenology, this work combines the knowledge of the physical, kinematic, and statistical properties of SAR imaging into a single unified Bayesian structure that simultaneously (a) estimates the nuisance parameters such as clutter distributions and antenna miscalibrations and (b) estimates the target signature required for detection/inference  of the target state.  Moreover, we provide a Monte Carlo estimate of the posterior distribution for the target state and nuisance parameters that infers the parameters of the model directly from the data, largely eliminating tuning of algorithm parameters.  We demonstrate that our algorithm competes at least as well on a synthetic dataset as state-of-the-art algorithms for estimating sparse signals.  Finally, performance analysis on a measured dataset demonstrates that the proposed algorithm is robust at detecting/estimating targets over a wide area and performs at least as well as popular algorithms for SAR moving target detection.

\end{abstract}

\section{Introduction}
\label{sar-sec:intro}
\ignore{In the previous chapters, tools were developed for sensor management and adaptive search for sparse moving targets.  This chapter provides an application for these tools, namely target detection and tracking with synthetic aperture radar (SAR) images.  The previous chapters relied on explicit characterizations of the uncertainty of target state estimation.  In Chapters \ref{chap:arap} and \ref{chap:darap}, one needed the posterior probabilities on target location and posterior estimates of the target amplitudes.  In Chapter \ref{chap:gum}, one needed estimates of the state estimation errors.  This chapter proposes inference algorithms and performance prediction (in terms of likelihood ratios and Cram\' er Rao lower bounds) that can be used explicitly for the algorithms in Chapters \ref{chap:arap} through \ref{chap:gum}.}

This work proposes algorithms for detecting and estimating targets in synthetic aperture radar (SAR) images.  The image formation process for SAR images is more complicated than that of standard electro-optical images.  Examples of these complexities include:
\begin{itemize}
\item SAR images have complex-valued rather than real-valued intensities, and the SAR phase information is of great importance for detection and estimation of  target states.  \cite{fienup2001detecting,deming2011along,deming2012three}.
\item SAR images are corrupted by spatiotemporally-varying antenna gain/phase patterns that often need to be estimated from homogeneous target-free data 
\cite{ranney2006signal,SoumM1997-UA-1}.
\item SAR images have spatially-varying clutter that can mask the target signature unless known a priori or properly estimated \cite{ender1999space}.
\item SAR images contain motion-induced displacement and diffusion of the target response \cite{jao2001theory,fienup2001detecting}.
\item SAR images include multiple error sources due to radar collection and physical properties of the reflectors, such as angular scintillation (a.k.a. glints) \cite{borden1983statistical} and speckle \cite{posner1993texture,raney1988spatial}.
\end{itemize}
Despite these complications, a great deal of structure exists in SAR images that can be leveraged to provide stronger SAR detection and tracking performance.  This includes (a) using the coherence between multiple channels of an along-track radar in order to remove the stationary background (a.k.a, `clutter'), (b) assuming that pixels within the image can be described by one (or a mixture) of a small number of object classes (e.g., buildings, vegetation, etc.), and (c) considering kinematic models for the target motion such as Markov smoothness priors.  From this structure in SAR imagery, one might consider models that assume that the clutter lies in a low-dimensional subspace that can be estimated directly from the data.  Indeed, recent work Borcea et al. \cite{borceaSARrpca} has shown that SAR signals can be represented as a composition of a low-rank component containing the clutter, a sparse component containing the target signatures, and additive noise.

In general, SAR images are formed by focusing the response of stationary objects to a single spatial location.  Moving targets, however, will cause phase errors in the standard formation of SAR images that cause displacement and defocusing effects.  Most methods designed to detect the target depend on either (a) exploiting the phase errors induced by the SAR image formation process for a single phase center system or (b) canceling the clutter background using a multiple phase center system.  In this chapter, we provide a rich model that can combine (and exploit) both sources of information in order to improve on both methodologies.

Fienup \cite{fienup2001detecting} provides an analysis of SAR phase errors induced by translational motions for single-look SAR imagery.  He shows that the major concerns are (a) azimuth translation errors from range-velocities, (b) azimuth smearing errors due to accelerations in range, and (c) azimuth smearing due to velocities in azimuth.  Fienup also provides an algorithm for detecting targets by their induced phase errors.  The algorithm is based on estimating the moving target's phase error, applying a focusing filter, and evaluating the sharpness ratio as a detection statistic.  Jao \cite{jao2001theory} shows that given both the radar trajectory and the target trajectory, it is possible to geometrically determine the location of the target signature in a reconstructed SAR image.  Although the radar trajectory is usually known with some accuracy, the target trajectory is unknown.  On the other hand, if the target is assumed to have no accelerations, Jao provides an efficient FFT-based method for refocusing a SAR image over a selection of range velocities.  Khwaja and Ma \cite{khwaja2011applications} provide a algorithm to exploit the sparsity of moving targets within SAR imagery; they propose a basis that is constructed from trajectories formed from all possible combinations of a set of velocities and positions.  To combat the computational complexity of searching through this dictionary, the authors use compressed sensing techniques.  Instead of searching over a dictionary of velocities, our work proposes to use a prior distribution on the target trajectory that can be provided a priori through road and traffic models or adaptively through observations of the scene over time. 

The process of removing the stationary background in order to detect moving targets is also known in the literature as `change detection' or `clutter suppresion'.  Gierull \cite{gierull2004statistical} provides a statistical analysis of the phase and magnitude of complex SAR images for two channels.  He shows that SAR images cannot be modeled as spatially-invariant Gaussian 
in many cases of interest, such as in urban environments, where the statistics vary spatially and may be modulated by random variations.  \ignore{The author provides probability density functions based on the p-distribution, as well as adaptive techniques for estimating the parameters of this distribution to be used in detecting slow-moving targets.}  In our work, we model the distributions of the clutter as spatially varying and model the random modulations directly. 

Ender \cite{ender1999space} applies space-time adaptive processing (STAP) to multiple-channel SAR imagery.  Similar to standard change detection algorithms such as displaced phase center array (DPCA) and along-track interferometry (ATI), STAP models the clutter as being embedded in a one-dimensional subspace.  However, STAP extends those algorithms to using $N>2$ channels, where a single channel is used to estimate the stationary background and the remaining $(N-1)$ channels are used to estimate the moving component.  However, STAP relies on estimating the complex-valued covariance matrix of the $N$-channel system, which in turn depends on the availability of homogeneous target-free secondary data.  

\ignoreForJournal{There are a multitude of algorithms for change detection that are based on multi-temporal SAR images rather than multi-channel data.  Bazi and Bruzzone \cite{bazi2005unsupervised} develop methods for multi-temporal change detection that use adaptive thresholds for declaring changes based on a theoretical analysis of a generalized Gaussian model.  Bovolo and Bruzonne \cite{bovolo2005detail} provide another algorithm for change detection that employs a wavelet-based multiple scale decomposition of multitemporal SAR images, with an adaptive scale driven fusion algorithm.}

Ranney and Soumekh \cite{ranney2006signal, SoumM1997-UA-1} develop methods for change detection from SAR images collected at two distinct times that are robust to errors in the SAR imaging process.  They address error sources including inaccurate position information, varying antenna gains, and autofocus errors.  They propose that the stationary components of multi-temporal SAR images can be related by a spatially-varying 2-dimensional filter.  To make the change detection algorithm numerically practical, the authors propose that this filter can be well-approximated by a spatially invariant response within small subregions about any pixel in the image.  This thesis adopts this model for the case where there are no registration errors.  Under a Gaussian assumption for the measurement errors, it can be shown that the maximum likelihood estimate for the filter coefficients can be computed easily through simple least squares.  

Ground Moving Target Indication (GMTI) methods involve the processing of SAR imagery to detect and estimate moving targets.  Often clutter cancellation and change detection play a preprocessing role in these algorithms \cite{perry1999sar,zhu2011ground,Guo-gmti-via-multichannel-SAR-2011,guo2011adaptive}.  This chapter aims to combine properties of many of these algorithms into a unifying framework that simultaneously estimates the target signature and the nuisance parameters, such as clutter distributions and antenna calibrations.  

It should be noted that many of the previously discussed algorithms work well in certain situations, but do not provide estimates of their uncertainty that may be necessary for adaptive sensing, sensor management, or sensor fusion.  This chapter aims to bridge this gap by providing a Bayesian formulation that provides uncertainty distributions for the presence of the moving targets and their positions.  Under this Bayesian formulation, we can generate the posterior distribution of the target state(s) given the observations (i.e., the SAR images).

Recently, there has been great interest by Wright et al. \cite{wright2009robust}, Lin et al. \cite{lin2010augmented}, Candes et al. \cite{candes2011robust} and Ding et al. \cite{ding2011bayesian} in the so-called robust principal component analysis (RPCA) problem that decomposes high-dimensional signals as
\begin{equation}
\label{sar-eq:rpca-equation}
{\bm I} = {\bm L} + {\bm S} + {\bm E},
\end{equation}
where  ${\bm I}\in\mathbb{R}^{N\times M}$ is an observed high dimensional signal, ${\bm L}\in\mathbb{R}^{N\times M}$ is a low-rank matrix with rank $r\ll NM$, ${\bm S}\in\mathbb{R}^{N\times M}$ is a sparse component, and  ${\bm E}\in\mathbb{R}^{N\times M}$ is dense low-amplitude noise.  In \cite{wright2009robust,lin2010augmented,candes2011robust}, inference in this model is done by optimizing a cost function of the form
\begin{equation}
\arg\min\limits_{\bm{L,S}} \norm{\bm L}_* + \gamma\norm{\bm S}_1 + (2\mu)^{-1}\norm{\bm{I-L-S}}_F
\end{equation}
where the last term is sometimes replaced by the constraint ${\bm I} = {\bm L}+{\bm S}$.  One major drawback of these methods involves finding the algorithm parameters (e.g.,  tolerance levels or choices of $\gamma,\mu$), which may depend on the given signal.  Moreover, it has been demonstrated that the performance of these algorithms can depend strongly on these parameters. 

Bayesian methods by Ding et al. \cite{ding2011bayesian} have been proposed that simultaneously learn the noise statistics and infer the low-rank and sparse components.  Moreover, they show that their method can be generalized to richer models, e.g. Markov dependencies on the target locations.  Additionally, these Bayesian inferences provide a characterization of the uncertainty of the outputs through a Markov Chain Monte Carlo (MCMC) estimate of the posterior distribution.

The work by Ding et al. \cite{ding2011bayesian} is based on a general Bayesian framework \cite{tipping2001sparse} by Tipping for obtaining \emph{sparse} solutions to regression and classification problems.  Tipping's framework uses simple distributions (e.g., those belonging to the exponential class) that can be described by few parameters, known as hyperparameters.  Moreover, Tipping considers a \emph{hierarchy} where the hyperparameters themselves are assumed to have a known `hyperprior' distribution.  Often the prior and hyperprior distributions are chosen to be conjugate, so that inference is simple.  Tipping provides insight into choosing the hyperparameter distributions so as to be non-informative with respect to the prior.  This latter property is important in making it possible to implement inference algorithms with few tuning parameters.  Finally, Tipping provides a specialization to the `relevance vector machine' (RVM), which can be thought of as a Bayesian version of the support vector machine.  Wipf et al. \cite{Wipf-04-sparse-bayes-learn-base-selec} provides an interpretation of the RVM as the application of a variational approximation to estimating the true posterior distribution.  Wipf et al. explains the sparsity properties of the sparse Bayesian learning algorithms in a rigorous manner.  Additionally, it also provides connections with other popular work in sparse problems, such as the FOCUSS and basis pursuit algorithms.

We adopt this hierarchical Bayesian model to SAR images.  This requires the following non-trivial extensions: (a) we consider complex-valued data rather than real-valued intensity images; (b) we model correlated noise sources based on physical knowledge of SAR phase history collection and image formation; (c) we relax the assumption of a low-rank background component by assuming that the background component lies in a low-dimensional subspace; and (d), we directly model SAR phenomenology by including terms for glints, speckle contributions, antenna gain patterns, and target kinematics.  Moreover, we demonstrate the performance of the proposed algorithm on both simulated and measured datasets, showing competitive or better performance in a variety of situations.  \perfpredswitch{}{Finally, we show that the output of the Bayesian model can be used for performance prediction for future passes of the radar.}

The rest of the paper is organized as follows: Notation is given in Section \ref{sar-sec:notation} and the image model is provided in \ref{sar-sec:image-model}.  Markov, spatial, and/or target kinematic extensions are discussed in Section \ref{sar-sec:markov-spatial-kinematic}.  The inference algorithm is given in Section \ref{sar-sec:inference}.  \perfpredswitch{}{ Section \ref{sar-sec:performance-prediction} provides theory for performance prediction using the output of the Bayesian inference.}  Performance is analyzed over both simulated and measured datasets in Section \ref{sar-sec:performance-analysis}.  We conclude and point to future work in Section \ref{sar-sec:conclusion}.  \forMastersThesis{Finally, in Section \ref{sar-sec:statistical-component}, we provide a discussion of the statistical components used in this work.}


\section{Notation}
\label{sar-sec:notation}
Available is a set of SAR images of a region formed from multiple passes of an along-track radar platform with multiple antennas (i.e., phase centers.)  Moreover, images are formed over distinct azimuth angle ranges that can be indexed by the frame number, $f$.  Table \ref{sar-table:index-variable-descriptions}
provides the indexing scheme used throughout this chapter in order to distinguish between images from various antennas, frames, and/or passes.  Table \ref{sar-table:data-conventions} provides a list of indexing conventions used to denote collections of variables. 

\renewcommand{\arraystretch}{1.25}
\begin{table}[thp]
\caption{Index variable names used in paper}
\label{sar-table:index-variable-descriptions}
\centering
\begin{tabular}{|l|c|c|}
\hline
{Index Description} & {Index Variable} &{Range}\\
\hline\hline
{Antenna (channel)} & $k$ & $1,2,\dots,K$\\
\hline
{Frame (azimuth range)} & $f$ & $1,2,\dots,F$\\
\hline
{Pass} & $i$ & $1,2,\dots,N$\\
\hline
{Pixel} & $p$ & $1,2,\dots,P$\\
\hline
\end{tabular}
\end{table}
\renewcommand{\arraystretch}{1.0}

\renewcommand{\arraystretch}{1.25}
\begin{table}[thp]
\caption{Our data indexing conventions}
\label{sar-table:data-conventions}
\centering
\begin{tabular}{|c|c|c|}
\hline
{Variable} & {Convention} &{Description}\\
\hline\hline
\multirow{2}{*}{$i_{k,f,i}^{(p)}$} & \multirow{2}{*}{Standard} & Value at pixel $p$, antenna $k$,\\
& & and frame $f$, pass $i$\\
\hline
\multirow{2}{*}{$\underline{i}_{f,i}^{(p)}$} & \multirow{2}{*}{Underline} & {Values at pixel $p$, frame $f$,}\\
& & and pass $i$ over all antennas\\
\hline
\multirow{2}{*}{${\bm i}_{f,1:N}^{(p)}$} & Lower-case, & Values at pixel $p$ and frame $f$\\
& Boldface & over all antennas and passes\\
\hline
\multirow{2}{*}{${\bm I}_{f,i}$} & Upper-case & Values over all pixels and\\
& Boldface &antennas at frame $f$ and pass $i$\\
\hline
\multirow{2}{*}{${\bm I}$} & Upper-case, & Values over all pixels, antennas,\\
& Boldface, No Indices& frames, and passes\\
\hline
\end{tabular}
\end{table}
\renewcommand{\arraystretch}{1.0}

We model the quadrature components of the SAR images with the complex-normal distribution, where we use the notation
\begin{equation}
\underline{w}\sim \mathcal{CN}\left(0,{\bm \Gamma}\right)
\end{equation}
where $\mathcal{CN}(\mu,{\bm \Gamma})$ represents the complex-Normal distribution with mean $\mu$ and complex covariance matrix ${\bm \Gamma}$, and $\vec{w}$ is random vector of $K$ complex-values (from each of $K$ antennas.)

\section{SAR image model}
\label{sar-sec:image-model}
\begin{figure*}[tp]
\label{sar-fig:graphical-model}
\begin{center}
\includegraphics[width=0.8\textwidth]{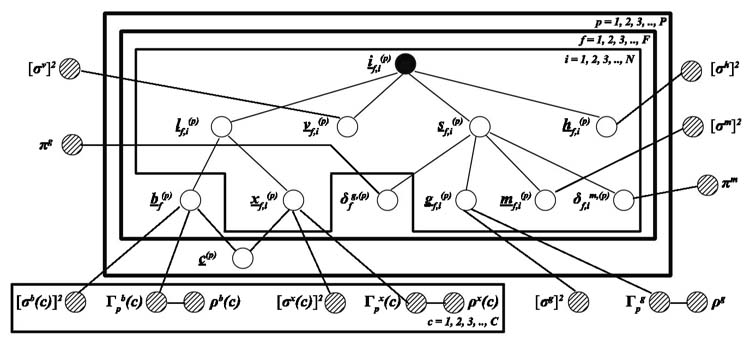}
\end{center}
\caption{This figure provides a graphical representation of the proposed SAR image model.  The dark circle represents the observed random variable.  The unshaded circles represent the basic parameters of the model, while the dashed circles represent hyperparameters that are also modeled as random variables.}
\end{figure*}

We propose a decomposition of SAR images at each frame $f$ and pass $i$ as follows
\begin{equation}
\label{sar-eq:image-decomposition-model}
\sarImat = \sarHmat \circ \left(\sarLmat + \sarSmat + \sarVmat\right),
\end{equation}
where $\sarHmat$ is a spatiotemporally-varying filter that accounts for antenna calibration errors, $\sarLmat$ is a low-dimensional representation of the background clutter, $\sarSmat$ is a sparse component that contains the targets of interest, $\sarVmat$ is zero-mean additive noise, and $\circ$ denotes the Hadamard (element-wise) product.  Each of these components belongs to the space $\mathbb{C}^{P\times K}$.  The remainder of this section discusses the model in detail.  Figure \ref{sar-fig:graphical-model} shows a graphical representation of the model.

\subsection{Low-dimensional component, $\sarLmat$}
We propose a decomposition of the low-rank component as
\begin{equation}
\label{sar-eq:background-decomposition-model}
\sarLmat = \sarBmat + \sarXmat,
\end{equation}
where $\sarBmat$ is the inherent background that is identical over all passes, $\sarXmat$ is the speckle noise component that arises from coherent imaging in SAR. Posner \cite{posner1993texture} and Raney \cite{raney1988spatial} describe speckle noise, which tends to be spatially correlated depending on the texture of the surrounding pixels.

The quadrature components of radar channels are often modeled as zero-mean Gaussian processes, though Gierull \cite{gierull2004statistical} demonstrates that for heterogeneous clutter (such as in urban scenes), one must consider spatially varying models.  To account for this spatial variation, this model assumes that 
each background pixel can be defined by one of $J$ classes that may be representative of roads, vegetation, or buildings within the scene.  Our model is low-dimensional since $J \ll P$, where $P$ is the number of pixels in the measured images.  We put a multinomial model on each object class 
\begin{equation}
\underline{c}^{(p)} = \set{c_j^{(p)}}_{j=1}^J \sim \mathrm{Multinomial}(1; q_1,q_2,\dots,q_J)
\end{equation}
where $q_j$ is the prior probability of the $j$-th object class.  Then the class assignment $C^{(p)}$ is the single location in $\underline{c}$ with value equal to one.  We use a hidden Markov model dependency that reflects that neighboring pixels are likely to have the same class.  The class $C^{(p)}$ defines the distribution of the pixel $p$, where we specifically model the background and speckle components respectively as complex-normal distributed:
\begin{equation}
\sarBv\sim \mathcal{CN}\left(0,{\bm \Gamma}_B^{C^{(p)}}\right),\quad\sarXv\sim \mathcal{CN}\left(0,{\bm \Gamma}_X^{C^{(p)}}\right)
\end{equation}
Note that the class type specifies the distribution of the pixels and each vector of $K$ values (e.g. background $\sarBv$ or speckle $\sarXv$) is drawn independently from that distribution.

\subsection{Sparse component, $\sarSmat$}
The sparse component is modeled as
\begin{equation}
\label{sar-eq:foreground-decomposition-model}
\sarSmat = \left(\sarDmatG\otimes{\bm 1}_K^T\right)\circ\sarGmat+\left(\sarDmatM\otimes{\bm 1}_K^T\right)\circ\sarMmat,
\end{equation}
where $\sarGmat\in\mathbb{C}^{P\times K}$ is the specular noise (glints) component with associated indicator variables $\sarDmatG\in\set{0,1}^P$, $\sarMmat\in\mathbb{C}^{P\times K}$ is the (moving) target component with associated indicator variables $\sarDmatM\in\set{0,1}^P$, ${\bm 1}_K$ is the all ones vector of size $K\times 1$, and $\otimes$ is the Kronecker product.  Note that this shared sparsity model assumes that the glint/target components are present in one antenna if and only if they are present in the other antennas.  Moreover, glints are known to have a large angular dependence, in the sense that the intensity of the glint dominates in only a few azimuth angles\ignoreForJournal{but is present from pass to pass as described by Borden \cite{borden1983statistical}}.  Thus, the indicators for glints do not depend on the pass index $i$.  Once again, we assume that the glints and target components are zero-mean complex-normal distributed with covariances ${\bm \Gamma}_G$ and ${\bm \Gamma}_M$, respectively.

The indicator variable $\delta^{z,(p)}$ at pixel $p$ where $z$ is representative of either $g$ or $m$ is modeled as
\begin{align}
\delta^{z,(p)} &\sim \mathrm{Bernoulli}(\pi^{z,(p)}),\\
\pi^{z,(p)} &\sim \mathrm{Beta}(a_\pi,b_\pi)
\end{align}
A sparseness prior is obtained by setting $a_\pi/[a_\pi+b_\pi]\ll 1$.  Alternatively, we can introduce additional structure in our
model by letting $a_\pi$ and $b_\pi$ depend on previous frames
(temporally) and/or neighboring pixels (spatially). This is
particularly useful for detecting multi-pixel targets that move
smoothly through a scene.  Section \ref{sar-sec:markov-spatial-kinematic} discusses this modification in greater detail.

\subsection{Distribution of quadrature components}
Many SAR detection algorithms rely on the ability to separate the target from the background clutter by assuming that the clutter lies in a low-dimensional subspace of the data. Consider a random vector of complex variables $\underline{w} \sim \mathcal{CN}\left(0,{\bm \Gamma}\right)$ where $w$ is representative of $b$, $x$, $g$ or $m$.  Under the assumptions that (a) the quadrature components of each antenna are zero-mean normal with variance $\sigma^2$ and (b) the correlation among components $w_{m}$ and $w_{n}$ is given by $\rho e^{-j\phi_{mn}}$, then ${\bm \Gamma}$ can be shown to have the form 
\begin{equation}
\label{sar-eq:complex-covariance-specific-model}
\begin{split}
{\bm \Gamma}&= \sigma^2 \left[ \begin{array}{*{20}c}
   {1} & {\rho e^{-j\phi_{12}}} & \cdots & {\rho e^{-j\phi_{1K}}}  \\
   {\rho e^{j\phi_{12}}} & {1} & \cdots & {\rho e^{-j\phi_{2K}}}  \\
   {\vdots} & {\vdots} & \ddots & \vdots  \\
   {\rho e^{j\phi_{1K}}} & {\rho e^{j\phi_{2K}}} & \cdots & {1}  \\
\end{array} \right],
\end{split}
\end{equation}
\ignore{
Note that the sample covariance matrix $\hat{{\bm \Gamma}}$, calculated from $Q$ independent samples of equation (\ref{sar-eq:complex-normal}) is known to be inverse-Wishart distributed with parameter ${\bm \Lambda} = (Q-K-1)E[{\bm \Gamma}]$.  Let ${\bm \Lambda}_0 = {\bm \Lambda}/(Q-K-1)$.  Then ${\bm \Lambda}_0$ has the form:
\begin{equation}
\label{sar-eq:complex-covariance-specific-model}
{\bm \Lambda}_0= \sigma^2 \left[ \begin{array}{*{20}c}
   {1} & {\rho e^{-j\phi_{12}}} & \cdots & {\rho e^{-j\phi_{1K}}}  \\
   {\rho e^{j\phi_{12}}} & {1} & \cdots & {\rho e^{-j\phi_{2K}}}  \\
   {\vdots} & {\vdots} & \ddots & \vdots  \\
   {\rho e^{j\phi_{1K}}} & {\rho e^{j\phi_{2K}}} & \cdots & {1}  \\
\end{array} \right],
\end{equation}
}
where $\sigma^2$ is the channel variance, $\rho$ is the coherence between antennas, and $\set{\phi_{nm}}_{n,m}$ are the interferometric phase differences between the antennas\footnote{A more general model could account for different channel variance and coherence values, but since we use the calibration constants $\sarHmat$ to equalize the channels, the effect was
seen to be relatively insignificant.}.  In an idealized model with a single point target, the interferometric phases $\phi_{mn}$ can be shown to be proportional to the target radial velocity.  In images containing only stationary targets (i.e., the background components), the covariance matrix has a simpler form:
\begin{equation}
\label{sar-eq:complex-covariance-background-model}
{\bm \Gamma}_{background} = \sigma^2\left((1-\rho)\eyeKK + \rho \oneK\oneK^T\right)
\end{equation}
where $\eyeKK$ is the $K\times K$ identity matrix and $\oneK$ is the all-ones vector of length $K$.

It should be noted that the covariance matrix in equation (\ref{sar-eq:complex-covariance-specific-model}) is related directly to some common methods for change detection in SAR imagery.  In particular, consider the two antenna case ($K=2$).  Along-track interferometry (ATI) thresholds the phase $\phi_{12}$ in order to detect moving targets which have non-zero phases.  Moreover, one can easily show that the eigendecomposition of ${\bm \Gamma}$ leads to eigenvalues $\lambda$ and eigenvectors $\nu$:
\begin{align}
\lambda({\bm \Gamma}) &= \set{2\sigma^2(1+\rho), 2\sigma^2(1-\rho)}\\
\nu({\bm \Gamma}) &= \set{\left[\begin{array}{c} 1\\e^{-j\phi_{12}}\end{array}\right], \left[\begin{array}{c} 1\\-e^{-j\phi_{12}}\end{array}\right]}.
\end{align}
Displaced phase center array (DPCA) processing thresholds the difference between the two channels.  Indeed, for small phases, the second eigenvector of ${\bm \Gamma}$ reduces to $[1;\ -1]^T$.  Thus DPCA can be interpreted as a projection onto the eigenvector of ${\bm \Gamma}$.  Deming \cite{deming2011along} shows that ATI performs well when canceling bright clutter (i.e., high $\sigma^2$ and $\rho\approx 1$), while DPCA performs well for canceling dim clutter (i.e., small $\sigma^2$ and $\rho\approx 0$.)  In our work, we combine the discriminating power of both DPCA and ATI by modeling the covariance matrices directly.  Ender \cite{ender1999space} provides space-time adaptive processing (STAP), where optimal detection schemes for moving targets are  based on the estimation of ${\bm \Gamma}$.  However, the performance of STAP depends on the availability of target-free homogeneously distributed measurements in order to estimate ${\bm \Gamma}$ effectively.  In this chapter, we simultaneously estimate the covariance matrices as well as the target contributions.  Thus, we demonstrate the capability to detect targets even in the presence of heterogeneous measurements.

In this thesis, the covariance matrix ${\bm \Gamma}$ is modeled as a random variable using a modified version of the Multivariate-Normal-Inverse-Wishart conjugate distributions.  In particular, we let
\begin{align}
\underline{w} &\sim {\mathcal{CN}}\left({\bm 0},{\sigma^2{\bm \Gamma}_\rho}\right)\\
\label{sar-eq:invwishart}{\bm \Gamma}_\rho &\sim \mathrm{InvWishart}\left(a_{\Gamma}((1-\rho)\eyeKK + \rho\oneK\oneK^T), \nu_\Gamma\right)\\
\sigma^2 &\sim \mathrm{InvGamma}(a_\sigma,b_\sigma)\\
\rho &\sim \mathrm{Beta}(a_\rho,b_\rho)
\end{align}
where $a_\sigma=b_\sigma=10^{-6}$ as suggested by Tipping\cite{tipping2001sparse} to promote non-informative priors,  $(a_\rho,b_\rho)$ are chosen so that $\rho\approx 1$ to ensure a high coherence among the background components, $\nu_\Gamma$ is a parameter that controls how strongly to weight the prior covariance matrix, and $a_\Gamma$ is chosen so that $E[{\bm \Gamma}_\rho] = (1-\rho)\eyeKK + \rho\oneK\oneK^T$.  In this thesis, we choose $\nu_\Gamma$ to be large to reflect our belief that $\sigma^2{\bm \Gamma}_\rho$ should be close to equation (\ref{sar-eq:complex-covariance-specific-model}).  Note that this model separates the learning of the channel variance $\sigma^2$, which we have no a priori knowledge about, from the learning of the correlation structure ${\bm \Gamma}_\rho$.

\subsection{Calibration filter, $\sarHmat$}
The calibration constants are assumed to be constant within small spatial regions $p \in Z_g$, though they may vary as a function of antenna, frame, or pass. In particular, we let
\begin{align}
h_{k,f,i}^{(p)} &= z_{k,f,i}(g), \forall p\in Z_g,\\
z_{k,f,i}(g) &\sim \mathcal{CN}(1, (\sigma^H)^2)
\end{align}
where we note that if $(\sigma^H)^2$ is large, then maximum likelihood inference in this case yields the least-squares solution.\ignore{  Thus, to save computational time, we may consider using LS instead of random sampling to estimate the calibration coefficients.}

\ignoreForJournal{
\subsection{Summary of SAR Image Model}

Tables \ref{sar-table:distributional-models} provides a summary of the distributions for the proposed decomposition of SAR images.  The table also provides a characterization of spatial (across pixels) and temporal (across frames and passes) dependencies.  For example, background and speckle components have distributions characterized by their class $j$.  Thus, all pixels with class $j$ belong to a subset $Q_j \subset \set{1,2,\dots,P}$.  In contrast, the distribution of moving targets is assumed to be identical across all pixels, yet the distribution of their indicators varies for each pixel, frame, and pass.

Tables \ref{sar-table:hyperparameters-distributional-models-covariances} and \ref{sar-table:hyperparameters-distributional-models-other} provide a summary of the parameters of the distributions in Table \ref{sar-table:distributional-models}.  We provide the simple model for target and glint indicator probabilities that just assumes that they are sparse in the image.  We can introduce additional richness in the model by allowing the parameters $a_\pi$ and $b_\pi$ to vary over pixels, frames, and passes as described in Section \ref{sar-subsection:indicator-prob}.

\renewcommand{\arraystretch}{1.25}
\begin{sidewaystable}[thp]
\caption{Distributional models for each component in equations (\ref{sar-eq:image-decomposition-model}), (\ref{sar-eq:background-decomposition-model}), and (\ref{sar-eq:foreground-decomposition-model})}
\label{sar-table:distributional-models}
\centering
\begin{tabular}{|l|c|l|c|c|c|}
\hline
{Component} & {Variable} & {Distribution} & {Parameters} & {Spatial} & {Temporal}\\
\hline\hline
{Background} & $\sarBv$ & $\cn{0}{\GammaB(j)}$ & $\GammaB(j)=[\sigma^B(j)]^2{\bm \Gamma}_\rho^B(j)$ & $p \in Q_j$ & All $f$, All $i$ \\
\hline
{Speckle} & $\sarXv$ & $\cn{0}{\GammaX(j)}$ &  $\GammaX(j)=[\sigma^X(j)]^2{\bm \Gamma}_\rho^X(j)$ & $p \in Q_j$& All $f$, All $i$\\
\hline
{Glints} & $\sarGv$ & $\cn{0}{\GammaG}$ &  $\GammaG=[\sigma^G]^2{\bm \Gamma}_\rho^G$ & All $p$& All $f$, All $i$\\
\hline
{Moving targets}& $\sarMv$ & $\cn{0}{\GammaM}$ &  $\GammaM=[\sigma^M]^2\eyeKK$ & All $p$& All $f$, All $i$\\
\hline
{Additive noise}&$\sarV$ & $\cn{0}{\GammaV}$ & $\GammaV=[\sigma^V]^2\eyeKK$ &All $p$& All $f$, All $i$\\
\hline
{Glint indicator}&$\sarDG$ & $\mathrm{Bernoulli}\left(\pi_{f}^{G,(p)}\right)$ & $\pi_{f}^{G,(p)}$ & Each $p$& Each $f$, All $i$\\
\hline
{Target indicator}&$\sarDM$ & $\mathrm{Bernoulli}\left(\pi_{f,i}^{M,(p)}\right)$ & $\pi_{f,i}^{M,(p)}$ & Each $p$& Each $f$, Each $i$\\
\hline
{Class assignment}& $\underline{c}(p)$ & $\mathrm{Multinomial}(1;\underline{q})$ & $\underline{q}$ & Each $p$& All $f$, All $i$\\
\hline
{Calibration coefficients} & $\sarH = z_{k,f,i}(g)$ &$\cn{1}{[\sigma^H]^2}$ & $[\sigma^H]^2$ & $p \in Z_g$& Each $f$, Each $i$\\
\hline
\end{tabular}
\end{sidewaystable}
\renewcommand{\arraystretch}{1.0}

\renewcommand{\arraystretch}{1.25}
\begin{sidewaystable}[thp]
\begin{minipage}{\textwidth}
\caption{Distributional models for covariance parameters of distributions in Table \ref{sar-table:distributional-models}}
\label{sar-table:hyperparameters-distributional-models-covariances}
\begin{center}
\begin{tabular}{|l|c|l|c|c|c|}
\hline
{Component} & {Variable} & {Distribution} & {Parameters} & {Suggested Value} & {Region}\\
\hline\hline
{Background covariance} & $\GammaB(j)=[\sigma^B(j)]^2{\bm \Gamma}_\rho^B(j)$ & {} & {} & {} & \multirow{4}{*}{Each $Q_j$}\\
{\quad Variance} & $\left[\sigma^B(j)\right]^2$ & {Inv-Gamma} & {$a_\sigma,\ b_\sigma$} & {$10^{-6},10^{-6}$}& {}\\
{\quad Correlation matrix} &${\bm \Gamma}_\rho^B(j)$ & {Inv-Wishart} & $a_\Gamma,\nu_\Gamma$ & See note$^b$, $O(P)$ & {}\\
{\quad Coherence} & $\rho^B(j)$ & {Beta} & {$a_\rho,\ b_\rho$} & $a_\rho/(a_\rho+b_\rho)\approx 1$ & {}\\
\hline
{Speckle covariance} & $\GammaX(j)=[\sigma^X(j)]^2{\bm \Gamma}_\rho^X(j)$ & {} & {} & {}& \multirow{4}{*}{Each $Q_j$}\\
{\quad Variance} & $\left[\sigma^X(j)\right]^2$ & {Inv-Gamma} & {$a_\sigma,\ b_\sigma$} & {$10^{-6},10^{-6}$}& {}\\
{\quad Correlation matrix} &${\bm \Gamma}_\rho^X(j)$ & {Inv-Wishart} & $a_\Gamma,\nu_\Gamma$ & See note$^b$, $O(P)$ & {}\\
{\quad Coherence} & $\rho^X(j)$ & {Beta} & {$a_\rho,\ b_\rho$} & {$a_\rho/(a_\rho+b_\rho)\approx 1$}& {}\\
\hline
{Glint covariance} & $\GammaG=[\sigma^G]^2{\bm \Gamma}_\rho^G$ & {} & {} & {}& \multirow{4}{*}{All $p$}\\
{\quad Variance}&$\left[\sigma^G\right]^2$ & {Inv-Gamma} & {$a_\sigma,\ b_\sigma$} & {$10^{-6},10^{-6}$}& {}\\
{\quad Correlation matrix}&${\bm \Gamma}_\rho^G$ & {Inv-Wishart} & $a_\Gamma,\nu_\Gamma$ & See note$^b$, $O(P)$ & {}\\
{\quad Coherence}&$\rho^G$ & {Beta} & {$a_\rho,\ b_\rho$} & {$a_\rho/(a_\rho+b_\rho)\approx 1$}& {}\\
\hline
{Target covariance}&$\GammaM=[\sigma^M]^2\eyeKK$ & {} & {} & {}& \multirow{2}{*}{All $p$}\\
{\quad Variance} &$\left[\sigma^M\right]^2$ & {Inv-Gamma} & {$a_\sigma,\ b_\sigma$} & {$10^{-6},10^{-6}$}& {}\\
\hline
{Additive noise covariance} & $\GammaV=[\sigma^V]^2\eyeKK$ & {} & {} & {}& \multirow{2}{*}{All $p$}\\
{\quad Variance} &$\left[\sigma^V\right]^2$ & {Inv-Gamma} & {$a_\sigma,\ b_\sigma$} & {$10^{-6},10^{-6}$}& {}\\
\hline
\end{tabular}
\end{center}
\ignore{$^a$It is assumed that the correlation matrix for the stationary components ($b$, $x$, and $g$) are distributed as in equation (\ref{sar-eq:invwishart}) so that the mean the shape parameter is given by $a_\Gamma \left((1-\rho)\eyeKK + \rho\oneK\oneK^T\right)$.\\
$^b$ The parameter $a_\Gamma$ is chosen so that $E[{\bm \Gamma}_\rho]=(1-\rho)\eyeKK + \rho\oneK\oneK^T$.}
\end{minipage}
\end{sidewaystable}
\renewcommand{\arraystretch}{1.0}

\renewcommand{\arraystretch}{1.25}
\begin{sidewaystable}[thp]
\caption{Distributional models for other parameters of distributions in Table \ref{sar-table:distributional-models}}
\label{sar-table:hyperparameters-distributional-models-other}
\begin{center}
\begin{tabular}{|l|c|l|c|c|c|}
\hline
{Component} & {Variable} & {Distribution} & {Parameters} & {Suggested Value} & {Region}\\
\hline\hline
 {Variance} &$\left[\sigma^V\right]^2$ & {Inv-Gamma} & {$a_\sigma,\ b_\sigma$} & {$10^{-6},10^{-6}$}& {}\\
\hline
{Target indicator probability} &$\pi_{f,i}^M(p)$ & {Beta} & {$a_\pi, b_\pi$} & {$a_\pi/(a_\pi+b_\pi)\ll 1$}& {Each $p,f,i$}\\
\hline
{Glint indicator probability}&$\pi_f^Y(p)$ & {Beta} & {$a_\pi, b_\pi$} & {$a_\pi/(a_\pi+b_\pi)\ll 1$}& {Each $p,f$}\\
\hline
{Region type probabilities}&{$\underline{q}=\set{q_1,\dots,q_J}$} & {Dirichlet} & {$\set{e_j}_{j=1}^J$} & {$e_j=1/J$}& {{All $p$}}\\
\hline
{Calibration coefficient variance} & $(\sigma^H)^2$ & {Inv-Gamma} & {$a_\sigma,\ b_\sigma$} & {$10^{-6},10^{-6}$}& {All $p$}\\
\hline
\end{tabular}
\end{center}
\end{sidewaystable}
\renewcommand{\arraystretch}{1.0}
}

\section{Markov/spatial/kinematic models for the sparse component}
\label{sar-sec:markov-spatial-kinematic}
\subsection{Indicator probability models}
\label{sar-subsection:indicator-prob}
This model contains multiple indicator variables with prior probabilities distributed as $\mathrm{Beta}(a_\pi,b_\pi)$.  Moreover, sparsity is obtained when $a_\pi/[a_\pi+b_\pi]\ll 1$.  Alternatively, we can introduce additional structure in our model by letting $a_\pi$ and $b_\pi$ depend on previous frames (temporally) and/or neighboring pixels (spatially).  This is especially useful for detecting multi-pixel targets that move smoothly through a scene.  

Define $W^M(p, \sarDmatM)$ to be a function that maps the indicator variables $\sarDmatM$ to a real number.  For example, this may be the average number of non-zero indicators in the neighborhood of pixel $p$, or a weighted version that puts higher value on neighboring pixels.  For $f=1$, we let
\begin{equation}
 \left[ \begin{array}{*{20}c} a_{1,i}^M(p)\\ b_{1,i}^M(p)\end{array}\right]
=
\begin{cases}
[a_H\quad b_H]^T, & {W^M(p, {\bm \Delta}_{1,i}^M)>\varepsilon_{spatial}^M,}\\
[a_L\quad b_L]^T, & \mathrm{else,}
\end{cases}
\end{equation}
and for $f>1$
\begin{equation}
\left[ \begin{array}{*{20}c} a_{f,i}^M(p)\\ b_{f,i}^M(p)\end{array}\right]
=
\begin{cases}
[a_H\quad b_H]^T, & {W^M(p, {\bm \Delta}_{f,i}^M)>\varepsilon_{spatial}^M\ \mathrm{and}}\\
& {W^M(p, {\bm \Delta}_{f-1,i}^M)>\varepsilon_{temporal}^M,}\\
[a_L\quad b_L]^T, & \mathrm{else.}
\end{cases}
\end{equation}
In this chapter, we choose $(a_L,b_L,a_H,b_H)$ so that $a_L/(a_L+b+L)\ll 1$ and $a_H/(a_H+b+H)\gg 0$.  A similar model can be introduced for the probabilities of the glints.

\subsection{Target kinematic model}
In some applications, such as target tracking or sequential detection, we may have access to an estimate of the kinematic state of the target(s) of interest, such as position, velocity and acceleration.  This may be useful for predicting the location of the target at sequential frames.  For simplicity, consider a single target at time $\tau$ whose state $\xi(\tau)=(\vr(\tau),\vdotr(\tau))$ is known with standard errors ${\bm \Sigma}_\xi(\tau)$.  Note that the uncertainty model for $(\vr,\vdotr)$ may be (a) known a prior from road maps or traffic behavior patterns, or (b) learned adaptively using some signal processing algorithm such as the Kalman or particle filters.  

In standard SAR image formation, moving targets tend to appear displaced and defocused as described by Fienup \cite{fienup2001detecting} and Jao\cite{jao2001theory}.  Moreover, Jao showed that given the radar trajectory $(\vq,\vdotq)$ and the target trajectory $(\vr,\vdotr)$, one can predict the location of the target signature within the image ${\bm p}$ by solving a system of equations that equate Doppler shifts and ranges, respectively, at each pulse:
\begin{align}
\frac{d}{d\tau}\left[\norm{\vp-\vq(\tau)}_2 - \norm{\vr(\tau)-\vq(\tau)}_2\right]_{\vp=\vp^*}=0\\
\norm{\vp^*-\vq(\tau)}_2=\norm{\vr(\tau)-\vq(\tau)}_2,
\end{align}
which can be reduced to the simpler system of equations:
\begin{align}
\label{sar-eq:equal-doppler}
\vdotq(\tau)\cdot[\vp^*-\vq(\tau)] &= [\vdotr(\tau)-\vdotq(\tau)]\cdot[\vr(\tau)-\vq(\tau)]\\
\label{sar-eq:equal-range}
\norm{\vp^*-\vq(\tau)}_2&=\norm{\vr(\tau)-\vq(\tau)}_2
\end{align}
The probable locations of the target can be predicted by one of several methods, including:
\begin{itemize}
\item Monte Carlo estimation of the target posterior density.
\item Gaussian approximation using linearization or the unscented transformation to approximate the posterior density
\item Analytical approximation.
\end{itemize}
Given an estimate of the posterior density, we can modify the function  $W^M$ described in the previous section to include dependence on this kinematic information.  \ignore{Alternatively, we could include the target state directly in our image model in order to simultaneously detect and estimate targets.}  Details of the posterior density estimation are provided 
\journalTechSwitch{in the technical report \cite{Newstadt-12-sar-techreport}.}{in Appendix \ref{sar-appendix:signature-prediction}.}

\section{Inference}
\label{sar-sec:inference}
\floatname{algorithm}{Procedure}
\renewcommand{\algorithmicrequire}{\textbf{Input:}}
\renewcommand{\algorithmicensure}{\textbf{Output:}}
\algrenewcommand{\algorithmiccomment}[1]{\hfill //\emph{\small{#1}}}
\begin{figure}[t]
\fbox{
\begin{minipage}{0.95\columnwidth}
\begin{changemargin}{-.15cm}{0cm}
{\bf procedure} $\set{\bm \Theta}_{i=1:N_{samples}}=$ SARGibbs$({\bm \Theta}_0,{\bm I})$
\end{changemargin}
\begin{changemargin}{-.2cm}{0cm}
\begin{algorithmic}
\State $\bm \Theta\gets{\bm \Theta}_0$
\For{$iteration=1$ to $N_{burnin}+N_{samples}$}
	\State Sample $\sim f\left({\bm B},{\bm X},{\bm G},{\bm M},{\bm \Delta}^G,{\bm \Delta}^M|{\bm I,-}\right)$ \Comment{Base}
	\State Sample $\sim f\left({\bm H}|{\bm I,-}\right)$\Comment{Calibration filter}
	\State Sample $\sim f\left({\bm C}|{\bm I,-}\right)$\Comment{Class assignment}
	\State Sample $\sim f\left({\bm \eta}|{\bm I,-}\right)$\Comment{Hyper-parameters}
\State ${\bm \Theta}_{iteration-N_{burnin}}\gets{\bm \Theta}$ {\bf if} $iteration>N_{burnin}$
\EndFor
\end{algorithmic}
\end{changemargin}
\begin{changemargin}{-.15cm}{0cm}
{\bf end procedure}
\end{changemargin}
\end{minipage}
}
\caption{Gibbs Sampling Pseudocode}          
\label{sar-alg:gibbs-sampler}                           
\end{figure}

In the proposed hierarchical model, the distribution of hyper-parameters at the base layer are generally chosen to be conjugate to the distributions at the next layer.  This allows for efficient approximation methods for the posterior distribution in the sense that we can sample exactly from these distributions.  In particular, we use a Markov Chain Monte Carlo (MCMC) algorithm in the form of a Gibbs sampler to iteratively estimate the full joint posterior.  In MCMC, this distribution is approximated by drawing samples iteratively from the conditional distribution of each (random) model variable given the most recent estimate of the rest of the variables (which we denote by $-$).  Let ${\bm \Theta}=\set{{\bm B},{\bm X},{\bm G},{\bm M},{\bm \Delta}^G,{\bm \Delta}^M,{\bm H},{\bm C},{\bm \eta}}$ represent a current estimate of all of the model variables where ${\bm \eta}$ represents the set of all hyper-parameters.  Given measurements $\bm I$, the inference algorithm is given in Figure \ref{sar-alg:gibbs-sampler}.  Note that MCMC algorithms require a burn-in period after the Markov chain has become stable, where the duration of burn-in period depends on the problem. After this point, we collect $N_{samples}$ samples that represent the full joint distribution.  \journalTechSwitch{Full details of the sampling procedures are given in the technical report\cite{Newstadt-12-sar-techreport}.}  However, we point out a couple of important features here.  First, the sampling of the base model can be rewritten as
\begin{align}
\label{sar-eq:samplestep-basic}&f({\bm B},{\bm X},{\bm G},{\bm M},{\bm \Delta}^G,{\bm \Delta}^M|{\bm I,-})\\
\nonumber&= \prod\limits_{p,f} f(\sarBv,\sarXvi,\sarGvi,\sarMvi,\sarDG,\sarDMi|{\bm I},-)
\end{align}
\ignore{where we use the underline notation to denote the collection of variables from different antennas $\underline{z}^{(p)} = [z_{1}^{(p)}\ \dots\ z_{K}^{(p)}]^T$ and boldface notation to denote a collection over passes ${\bm z}_{1:N}^{(p)} = [\underline{z}_1^{(p)}\ \cdots \underline{z}_N^{(p)}]^T$. } The conditional independence among pixels and frames given the nuisance parameters allows us to easily parallelize the sampling procedure over the largest dimensions of the state.  Moreover, we can extend the parallelization to sampling independently over passes by separating the sampling of equation (\ref{sar-eq:samplestep-basic}) into two Gibbs steps from the densities:
\begin{align}
\label{sar-eq:samplestep-basic2}
&f(\sarBv,\sarXvi,\sarMvi,\sarDMi|{\bm I},-)\\
\nonumber&\qquad= f(\sarBv|{\bm I},\sarXvi,\sarMvi,\sarDMi,-)\\
\nonumber&\qquad\cdot\prod_{i} f(\sarXv,\sarMv|\sarDM,{\bm I},-)f(\sarDM|{\bm I},-)\\
\label{sar-eq:samplestep-basic3}&f(\sarGvi,\sarDG|{\bm I},-)\\
\nonumber&\qquad=f(\sarDG|{\bm I},-)\prod_{i} f(\sarGv|{\bm I},\sarDG,-)
\end{align}
In both of the sampling steps in equations (\ref{sar-eq:samplestep-basic2}) and (\ref{sar-eq:samplestep-basic3}), we have an exact inference algorithm over multivariate-Gaussian distributed variables and Bernoulli distributed variables.  This leads to faster convergence of the Markov chain and subsequently fewer burn-in samples.  The conditional density for the nuisance parameters ${\bm \eta}$ given the remainder variables can also be re-written to allow for efficient sampling.  In particular, due to conditional independence we have:
\begin{equation}
\begin{split}
f({\bm \eta}|{\bm I},-) = &f({\bm \Gamma}^M|{\bm M},{\bm \Delta}^M)f({\bm \pi}^M|{\bm M},{\bm \Delta}^M,{\bm \Gamma}^M)\\
\cdot &f({\bm \Gamma}^G|{\bm G},{\bm \Delta}^G)f({\bm \pi}^G|{\bm G},{\bm \Delta}^G,{\bm \Gamma}^G)\\
\cdot&\prod\limits_j f({\bm \Gamma}^B(j)|{\bm B},{\bm C}) f({\bm \Gamma}^X(j)|{\bm X},{\bm C})
\end{split}
\label{sar-eq:gibbs-nuisance}
\end{equation}
\ignore{
\begin{equation}
\begin{split}
f&({\bm \eta}|{\bm I},-) = f({\bm \Gamma}^M|{\bm M},{\bm \Delta}^M)f({\bm \pi}^M|{\bm M},{\bm \Delta}^M,{\bm \Gamma}^M)\\
&\cdot\prod\limits_j f({\bm \Gamma}^B(j)|{\bm B},{\bm C}) f({\bm \Gamma}^X(j)|{\bm X},{\bm C})\\
&\cdot f({\bm \pi}^G|{\bm G},{\bm \Delta}^G,\set{{\bm \Gamma}^G(j)}_j)\prod\limits_j f({\bm \Gamma}^G(j)|{\bm G},{\bm \Delta}^G,{\bm C})
\end{split}
\label{sar-eq:gibbs-nuisance}
\end{equation}
}
where ${\bm \Gamma}$ represents the parameters related to the covariance matrices (i.e., the variance $\sigma^2$, correlation structure ${\bm \Gamma}_\rho$, and the coherence $\rho$).  Once again, this decomposition allows for a sampling procedure that leads to faster convergence of the Gibbs sampler.  Moreover, the sampling procedures for  the individual densities in equation (\ref{sar-eq:gibbs-nuisance}) tend to require sufficient statistics that are of significantly smaller dimension and thus more desirable from a computational viewpoint.  For example, sampling of the covariance matrix ${\bm \Gamma}^M$ depends only on a $K\times K$ sample covariance matrix.  It should be noted that sampling of the covariance matrices requires additional effort in order to constrain its shape to that of equation (\ref{sar-eq:complex-covariance-specific-model}).  In particular, we use a Metropolis-Hastings step, which can be easily done by noting that the posterior density $f({\bm \Gamma}^W,\rho^W,(\sigma^2)^W|{\bm W})$ is proportional to an Inverse-Wishart distribution.  Details are provided \journalTechSwitch{in the technical report\cite{Newstadt-12-sar-techreport}.}{in Appendix \ref{sar-appendix:inference}.}

\perfpredswitch{}{\section{Performance prediction}
\label{sar-sec:performance-prediction}
\input{sar_PerformancePrediction}
}
\section{Performance analysis}
\label{sar-sec:performance-analysis}
\subsection{Simulation}
\renewcommand{\arraystretch}{1.25}
\begin{table}[tp]
\caption{Parameters of simulated dataset}
\label{sar-table:simulated-dataset-parameters}
\centering
\ignore{
\begin{tabular}{cc}
\firstrowfirstcell{Parameter} &\firstrowcell{Value} \\
\oddTableRowc{Pixels in image, $P$}{$P = 100\times 100$}\\
\evenTableRowc{Number of frames per pass, $F$}{$F=1$}\\
\oddTableRowc{\# of antennas, $K$}{$K=3$}\\
\evenTableRowc{\# of passes, $N$}{$N\in\set{5,10,20}$}\\
\oddTableRowc{\# of target pixels/image, $N_{targets}$}{$N_{targets}=20$}\\
\evenTableRowc{Clutter of background, $\rho$}{$\rho\in\set{0.9,0.99,0.999,0.9999}$}\\
\oddTableRowc{Variance of targets, $\sigma_{target}^2$}{$\sigma_{target}^2=1$}\\
\evenTableRowc{Variance of background}{Either $\sigma_{dim}^2=\sigma_{clutter}^2/100$}\\
\evenTableRowc{}{or $\sigma_{bright}^2=\sigma_{clutter}^2$}\\
\oddTableRowc{Signal-to-noise-plus clutter (SCNR)}{$\mathrm{SCNR}\trieq\frac{\sigma_{target}^2}{\sigma_{clutter}^2+\sigma_{noise}^2}$}\\
\oddTableRowc{}{$\in\set{0.1,0.5,1,2}$}\\
\end{tabular}
}
\begin{tabular}{|l|c|}
\hline
{\bf Parameter} &{\bf Value} \\
\hline
\hline
{Pixels in image, $P$} & {$P = 100\times 100$}\\\hline
{Number of frames per pass, $F$} &{$F=1$}\\\hline
{\# of antennas, $K$} &{$K=3$}\\\hline
{\# of passes, $N$} & {$N\in\set{5,10,20}$}\\\hline
{\# of target pixels/image, $N_{targets}$}& {$N_{targets}=20$}\\\hline
{Clutter of background, $\rho$} & {$\rho\in\set{0.9,0.99,0.999,0.9999}$}\\\hline
{Variance of targets, $\sigma_{target}^2$} & {$\sigma_{target}^2=1$}\\\hline
\multirow{2}{*}{Variance of background} & {Either $\sigma_{dim}^2=\sigma_{clutter}^2/100$}\\
{} & {or $\sigma_{bright}^2=\sigma_{clutter}^2$}\\\hline
\multirow{2}{*}{Signal-to-noise-plus clutter (SCNR)}  & {$\mathrm{SCNR}\trieq\frac{\sigma_{target}^2}{\sigma_{clutter}^2+\sigma_{noise}^2}$}\\
{} & {$\in\set{0.1,0.5,1,2}$}\\
\hline
\end{tabular}
\end{table}
\renewcommand{\arraystretch}{1.0}

\begin{figure}[t]
\centering
\includegraphics[width=0.95\columnwidth]{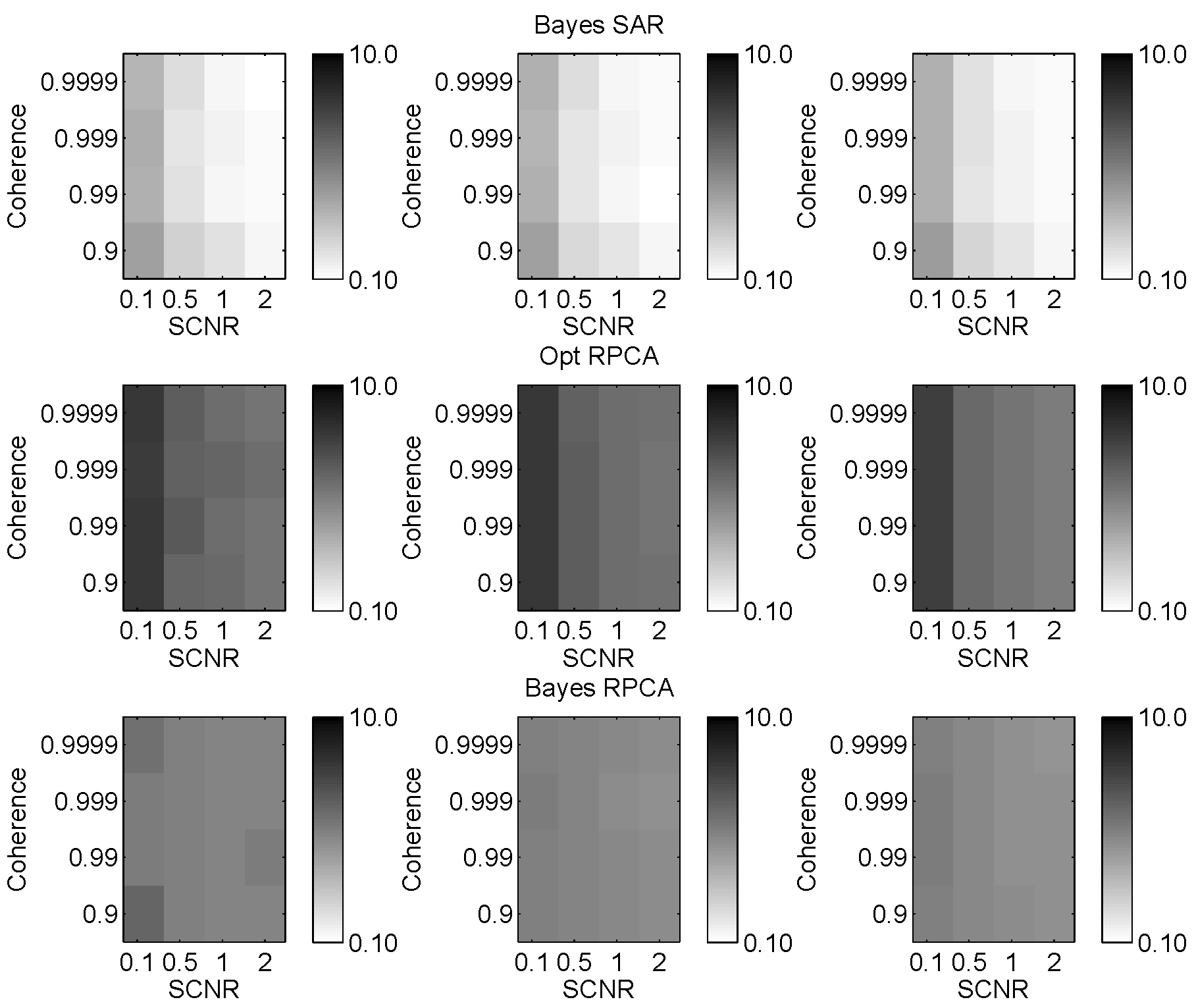}
\caption{
This figure compares the relative reconstruction error of the target component, $\frac{\norm{{\bm S}-\hat{\bm S}}_2}{\norm{\bm S}_2}$, as a function of algorithm, number of passes $N$, coherence of antennas $\rho$, and signal-to-clutter-plus-noise ratio (SCNR).  From top-to-bottom, the rows contains the output of the Bayes SAR algorithm, the optimization-based RPCA algorithm, and the Bayes RPCA algorithm.  From left-to-right, the columns show the output for $N=5$, $N=10$, and $N=20$ passes (with $F=1$ frames per pass).  The output is given by the median error over 20 trials on a simulated dataset.  It is seen that in all cases, the Bayes SAR method outperforms the RPCA algorithms.  Moreover, the Bayes SAR algorithm performs better if either coherence increases (i.e., better clutter cancellation) or the SCNR increases.  On the other hand, the performance of the RPCA algorithms does not improve with increased coherence, since these algorithms do not directly model this relationship.
}
\label{sar-fig:comparisons-rpca}
\end{figure}

\begin{figure}[t]
\centering
\journalTechSwitch{
\subfigure[${\bm L}+{\bm S}$\hspace{0.3in}]{
\includegraphics[width=.13\textwidth]{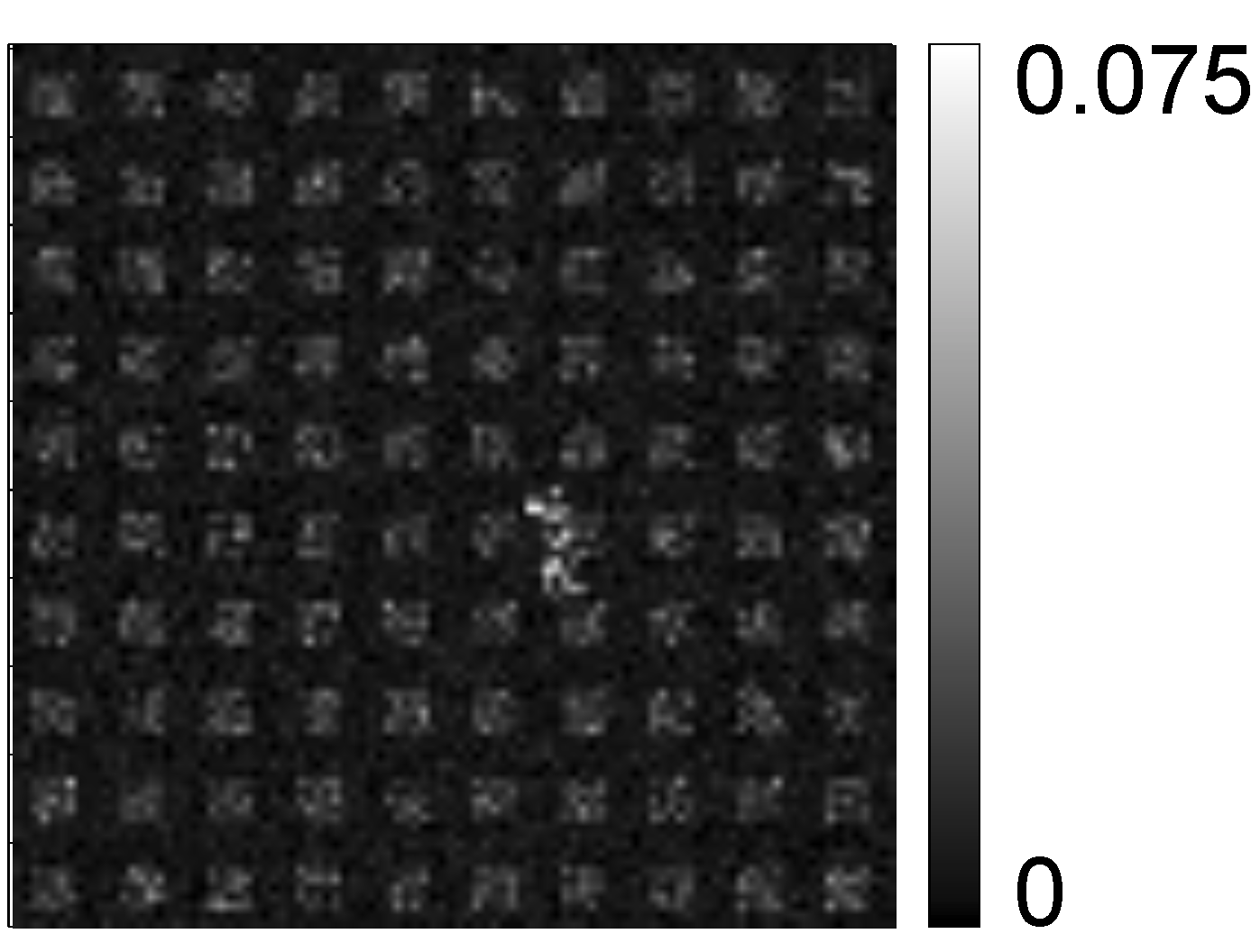}
}
\subfigure[${\bm L}$\hspace{0.3in}]{
\includegraphics[width=.13\textwidth]{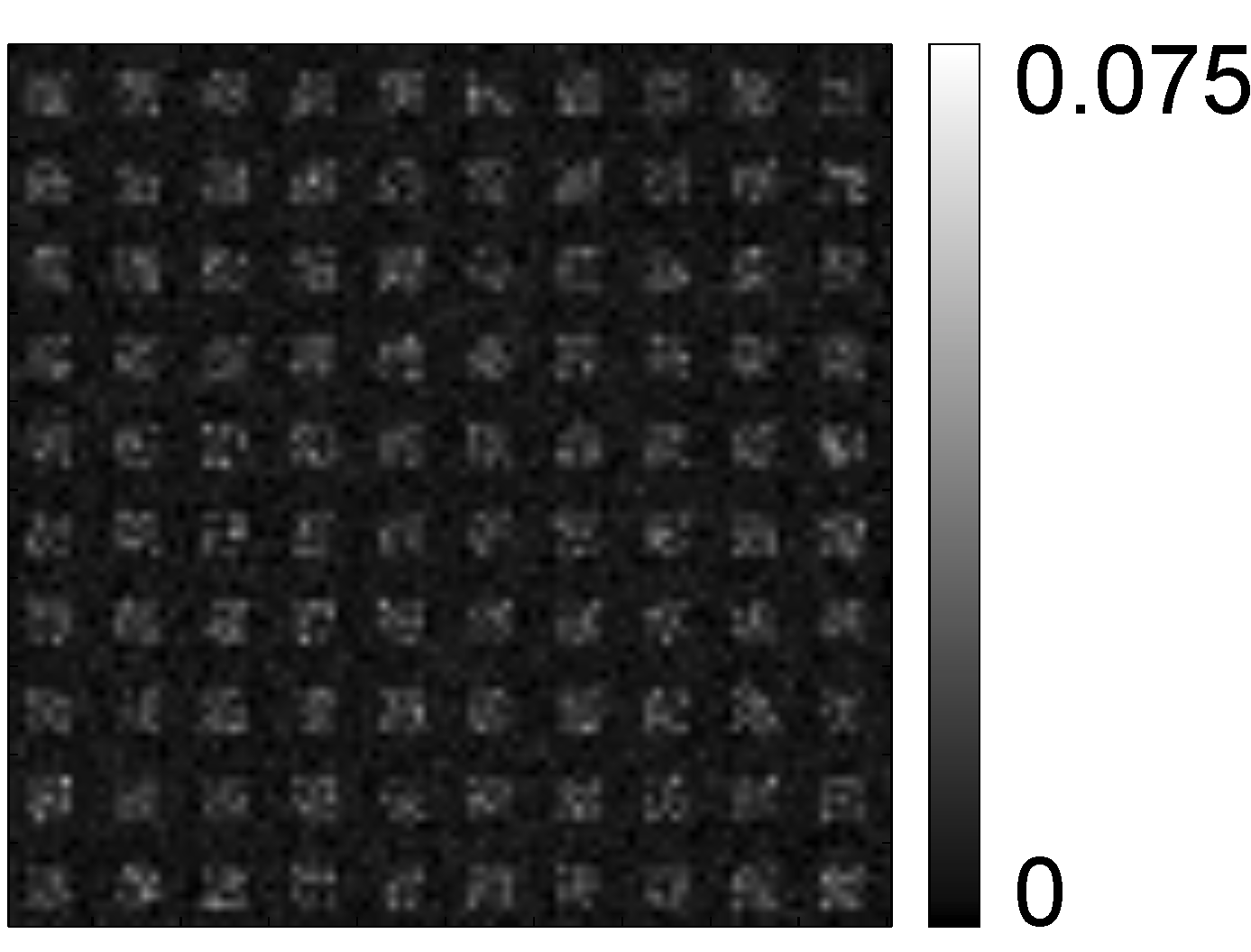}
}
\subfigure[${\bm S}$\hspace{0.3in}]{
\includegraphics[width=.13\textwidth]{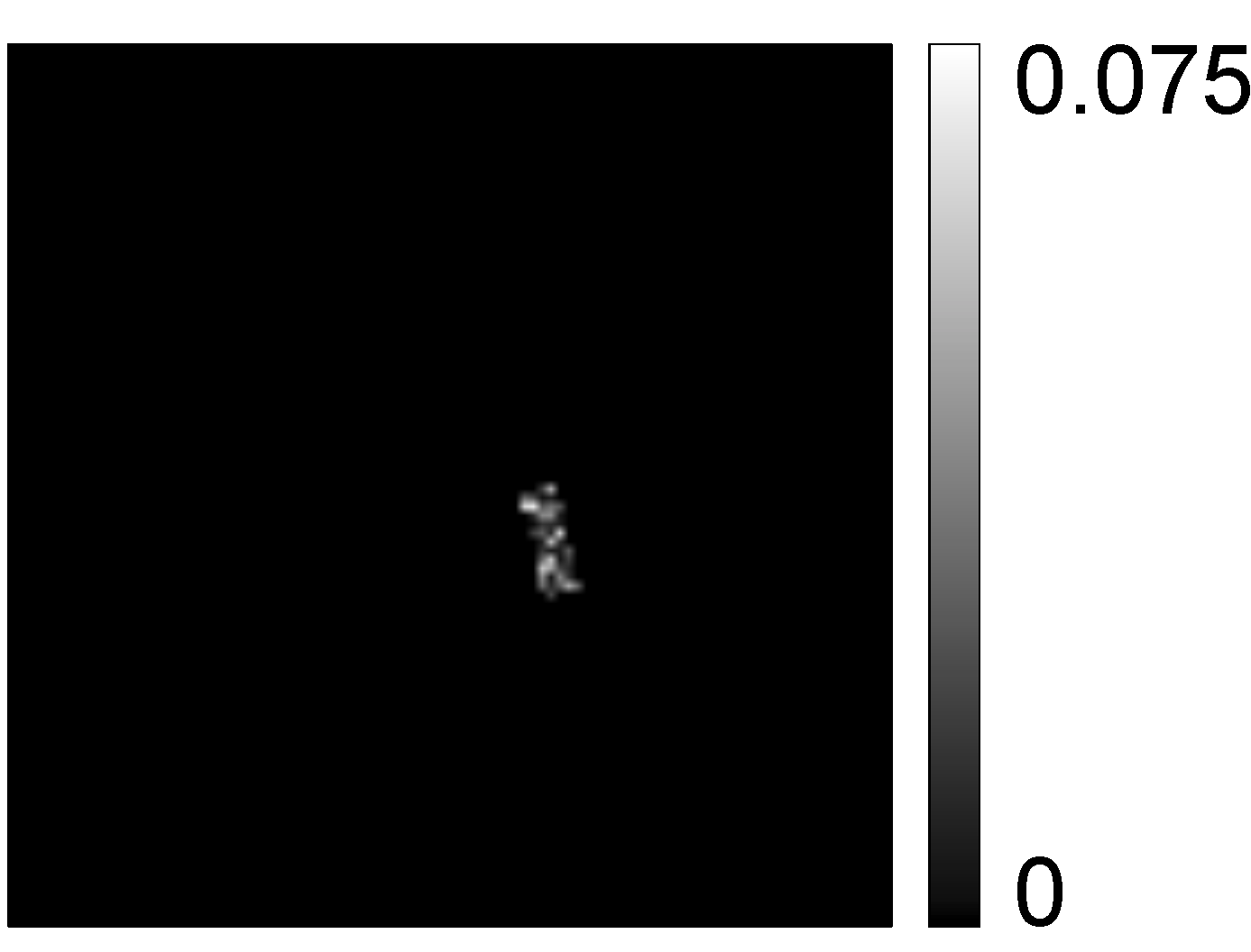}
}
}{
\subfigure[${\bm L}+{\bm S}$\hspace{0.3in}]{
\includegraphics[width=.3\textwidth]{SimulatedImage_ver2}
}
\subfigure[${\bm L}$\hspace{0.3in}]{
\includegraphics[width=.3\textwidth]{SimulatedBackground_ver2}
}
\subfigure[${\bm S}$\hspace{0.3in}]{
\includegraphics[width=.3\textwidth]{SimulatedTarget_ver2}
}
}
\caption{
This figure provides a sample image used in the simulated dataset for comparisons to RPCA methods, as well as its decomposition into low-dimensional background and sparse target components.  This low SCNR image is typical of measured SAR images.  Note that the target is randomly placed within the image for each of $N$ passes.  In some of these passes, the target is placed over low-amplitude clutter and can be easily detected.  In other passes, the target is placed over high-amplitude clutter, which reduces the capability to detect the target.
}
\label{sar-fig:sample-rpca}
\end{figure}

\renewcommand{\arraystretch}{1.25}
\begin{table}[thp]
\caption{Comparison of proposed method (Bayes SAR) to RPCA Methods with $N=20$, $F=1$, $K=3$.  Note that the Bayes SAR method performs about twice as well as either of the RPCA methods for all criteria.  The Bayes SAR method also produces a sparse result.}
\label{sar-table:comparisons-rpca}
\centering
\subtable[Bayes SAR]{
\begin{tabular}{|c|c|c|c|c|}
\hline
{SCNR} &{Coherence} 
&{$\frac{\norm{{\bm L}-\hat{\bm L}}_2}{\norm{\bm L}_2}$} &{$\frac{\norm{{\bm S}-\hat{\bm S}}_2}{\norm{\bm S}_2}$} &{$\frac{\norm{{\bm S}-\hat{\bm S}}_0}{\norm{\bm S}_0}$}\\
\hline\hline
{10\%}&$0.900$ & $0.057$ & $0.578$ & $0.550$\\\hline
{10\%}&$0.9999$ & $0.045$ & $0.419$ & $0.367$\\\hline
{100\%}&$0.900$ & $0.057$ & $0.155$ & $0.150$\\\hline
{100\%}&$0.9999$ & $0.052$ & $0.122$ & $0.096$\\\hline
{200\%}&$0.900$ & $ $0.057& $0.123$ & $0.137$\\\hline
{200\%}&$0.9999$ & $0.056$ & $0.114$ & $0.092$\\\hline
\end{tabular}
}
\subtable[Opt. RPCA]{
\begin{tabular}{|c|c|c|c|c|}
\hline
{SCNR} &{Coherence} 
&{$\frac{\norm{{\bm L}-\hat{\bm L}}_2}{\norm{\bm L}_2}$} &{$\frac{\norm{{\bm S}-\hat{\bm S}}_2}{\norm{\bm S}_2}$} &{$\frac{\norm{{\bm S}-\hat{\bm S}}_0}{\norm{\bm S}_0}$}\\
\hline\hline
{10\%}&$0.900$ & $0.111$ & $3.175$ & $111.026$\\\hline
{10\%}&$0.9999$ & $0.113$ & $3.237$ & $109.716$\\\hline
{100\%}&$0.900$ & $0.111$ & $1.189$ & $109.520$\\\hline
{100\%}&$0.9999$ & $0.110$ & $1.173$ & $108.203$\\\hline
{200\%}&$0.900$ & $ $0.112& $1.058$ & $111.120$\\\hline
{200\%}&$0.9999$ & $0.110$ & $1.035$ & $109.583$\\\hline
\end{tabular}
}
\subtable[Bayes RPCA]{
\begin{tabular}{|c|c|c|c|c|}
\hline
{SCNR} &{Coherence} 
&{$\frac{\norm{{\bm L}-\hat{\bm L}}_2}{\norm{\bm L}_2}$} &{$\frac{\norm{{\bm S}-\hat{\bm S}}_2}{\norm{\bm S}_2}$} &{$\frac{\norm{{\bm S}-\hat{\bm S}}_0}{\norm{\bm S}_0}$}\\
\hline\hline
{10\%}&$0.900$ & $0.117$ & $0.998$ & $3.761$\\\hline
{10\%}&$0.9999$ & $0.108$ & $0.990$ & $3.799$\\\hline
{100\%}&$0.900$ & $0.116$ & $0.764$ & $3.451$\\\hline
{100\%}&$0.9999$ & $0.117$ & $0.741$ & $3.494$\\\hline
{200\%}&$0.900$ & $ $0.125& $0.706$ & $3.665$\\\hline
{200\%}&$0.9999$ & $0.129$ & $0.692$ & $3.720$\\
\hline
\end{tabular}
}
\end{table}
\renewcommand{\arraystretch}{1.0}

We first demonstrate the performance of the proposed algorithm, which we refer to as the Bayes SAR algorithm, on a simulated dataset.  Images were created according to the model given in Section \ref{sar-sec:image-model} with parameters given in Table \ref{sar-table:simulated-dataset-parameters}.  The low-dimensional component was divided into one of two classes (`dim' or `bright'). Pixels were deterministically assigned to one of these classes to resemble a natural SAR image (see Figure \ref{sar-fig:sample-rpca}).  The sparse component included a randomly placed target with multiple-pixel extent. A spatiotemporally varying antenna gain filter was uniformly drawn at random on the range $[0,2\pi)$ for groups of pixels of size $25\times 25$.  Lastly, zero-mean IID noise was added with variance $\sigma_{noise}^2$.
\ignore{Each set of $K$ pixels values for the low-dimensional components are independently drawn at random from zero-mean complex-normal distributions with covariance matrix given by equation (\ref{sar-eq:complex-covariance-specific-model}) with variance given by $\sigma_{dim}^2$ or $\sigma_{bright}^2$, with coherence $\rho_{clutter}$, and with interferometric phases $\phi_{mn}=0$ for all $m,n$. A target object with multiple pixel extent (approximately $0.2\%\ P$ pixels) was randomly placed in the scene for each of $N$ passes with variance $\sigma_{target}^2$, coherence $\rho_{target}=0$, and randomly drawn interferometric phases $\phi_{mn}$.  A spatially-varying antenna phase pattern was uniformly drawn at random on the range $[0,2\pi)$ for groups of pixels of size $25\times 25$.  Lastly, zero-mean IID noise was added to the scene with variance $\sigma_{noise}^2=\sigma_{clutter}^2/500$.  We consider various values of $N$ (the number of passes), $\rho_{clutter}$, and the signal-to-clutter-plus-ratio (SCNR) defined as $\mathrm{SCNR}=\sigma_{target}^2/(\sigma_{clutter}^2+\sigma_{noise}^2)$.  Note that increasing any of these quantities should improve performance by either improving signal quality or allowing for better clutter cancellation.}

The Bayes SAR model is applied to infer the low-dimensional component $\sarLmat$ and sparse target component $\sarSmat$ with estimates denoted $\hat{\bm L}_{f,i}$ and $\hat{\bm S}_{f,i}$, respectively.  Hyperparameters of the model \ignore{$(a_\sigma,b_\sigma,a_\rho,b_\rho,a_H,b_H,a_L,b_L)$} are chosen according to the Section \ref{sar-sec:inference}.  Results are given by the mean of MCMC inference with 500 burn-in iterations followed by 100 collection samples.  We consider three metrics to evaluate the reconstruction errors: $\frac{\norm{{\bm L}-\hat{\bm L}}_2}{\norm{\bm L}_2}$, $\frac{\norm{{\bm S}-\hat{\bm S}}_2}{\norm{\bm S}_2}$, $\frac{\norm{{\bm S}-\hat{\bm S}}_2}{\norm{\bm S}_0}$, where the norm is taken over the vectorized quantities.\ignore{  Note that it might be possible to have good reconstruction $l_2$ errors for $\bm L$ and $\bm S$, but not be sparse as given by the $l_0$ norm.  This may negatively impact estimation of the target state.}  

In comparison to the Bayes SAR model, results are given for state-of-the-art algorithms for Robust Principal Component Analysis (RCPA): an optimization-based approach proposed by Wright et al. \cite{wright2009robust} and Candes et al. \cite{candes2011robust} and a Bayesian-based approach proposed by Ding et al. \cite{ding2011bayesian}\footnote{For the optimization-based approach, we used the exact\_alm\_rpca package (MATLAB)  by Lin et al. \cite{lin2010augmented}, downloaded from \url{http://watt.csl.illinois.edu/perceive/matrix-rank/home.html}.  For the Bayesian-based approach, we used the Bayesian robust PCA package, downloaded from \url{http://www.ece.duke.edu/~lihan/brpca_code/BRPCA.zip}.}.  \ignore{It should be noted that both of these algorithms were designed for real-valued data.  Thus, the input to the algorithms were the magnitude and phase components of the generated images.}   The optimization-based approach requires a tolerance parameter which is related to the noise level, as suggested by Ding et al. \cite{ding2011bayesian}.  We chose this parameter in order to have the smallest reconstruction errors.  The Bayesian method did not require tuning parameters, except for choosing the maximum rank of $\sarLmat$ which was set to 20.

Figure \ref{sar-fig:comparisons-rpca} compares the relative reconstruction error of the sparse (target) component, $\frac{\norm{{\bm S}-\hat{\bm S}}_2}{\norm{\bm S}_2}$, across all algorithms, number of passes $N$, coherence of antennas $\rho$, and SCNR.   In all cases, the Bayes SAR method outperforms the RPCA algorithms with improving performance if either coherence or SCNR increases.  \ignore{On the other hand, the performance of the RPCA algorithms does not improve with increasing coherence, since these algorithms do not directly model this relationship.}  Table \ref{sar-table:comparisons-rpca} provides additional numerical results for the case $N=20$.  The RCPA algorithms perform poorly in reconstructing the sparse component with relative errors near or greater than 1.  This reflects the fact that (a) these algorithms miss significant sources of information, such as the correlations among antennas and among quadrature components, and (b) $N=20$ may be too few samples to reliably estimate the principal components in these non-parametric models.  In measured SAR imagery, it might be unreasonable to expect $N\gg 20$ passes of the radar, which suggests that these RPCA algorithms will likely perform poorly on such signals.  In contrast, it is seen that the Bayes SAR method obtains low reconstruction errors for both low-dimensional and sparse components as either coherence or SCNR increase.

\subsection{Measured data}
In this section, we compare performance of the Bayes SAR approach using a set of measured data from the 2006 Gotcha SAR sensor collection.  In particular, images were formed from phase histories collected over a scene of size 375m by 1200m for $N=3$ passes and $K=3$ antennas.  Each image was created with a coherent processing time of 0.5 seconds with the addition of a Blackman-Harris window in the azimuth direction to reduce sidelobes.  Images were created at overlapping intervals spaced 0.25 seconds apart for a total of 18 seconds.  Note that the ability to take advantage of correlated images (as in this case) is one of the benefits of using the proposed model/inference algorithm.

\ignore{Performance in this section is illustrated visually to demonstrate the capabilities of the proposed model in a variety of situations.  However, in addition to the visual results, the dataset we considered also contained a few GPS-truthed vehicles for which we can make quantitative comparisons.  In particular, we use the GPS-given target state and equations (\ref{sar-eq:equal-doppler}) and (\ref{sar-eq:equal-range}) in order to determine (a) the `true' location of the target within the formed SAR image, and (b) the target's radial velocity which is known to be proportional to the measured interferometric phase of the target pixels in an along-track system.}  

We consider three alternative approaches in comparison to the Bayes SAR approach: (1) displaced-phase center array (DPCA) processing, (2) along-track interferometry (ATI), and (3) a mixture of DPCA/ATI. \ignore{DPCA thresholds the differences between images formed from two antennas in an along-track system.  In an ideal noise-free setting, only moving targets will have noticeable differences from antenna to antenna.  In practice, however, small variations in high-clutter regions will cause significant false alarms in DPCA.  In contrast, ATI thresholds the interferometric phase differences in an along-track system.  Since bright-clutter tends to be highly correlated, ATI is able to effectively cancel clutter and leave only the targets.  In practice, dim-clutter regions will cause false alarms in ATI, since their phases tend to be more adversely affected by noise.  To improve upon both of these methods, Deming \cite{deming2011along,deming2012three} proposes algorithms to use ATI to cancel bright clutter and DPCA to cancel dim clutter, leaving only the targets themselves.}  Note that all variants of ATI/DPCA depend on the chosen thresholds for phase/magnitude, respectively.

\subsubsection{Comparisons to DPCA/ATI}
\begin{figure*}[thp]
\centering
\subfigure[Magnitude (original)\hspace{0.3in}]{
\includegraphics[height=.18\textwidth]{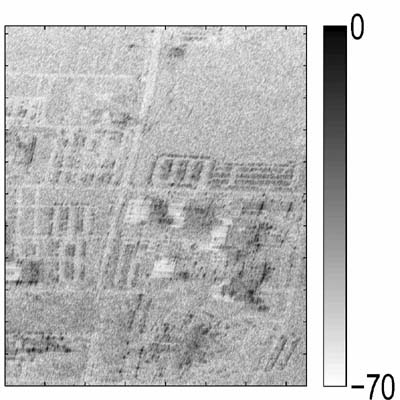}}
\subfigure[Phase (original)\hspace{0.3in}] {\includegraphics[height=.18\textwidth]{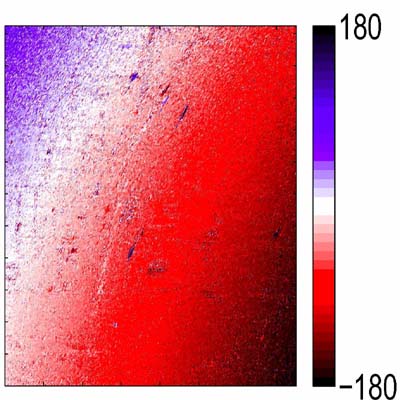}}
\subfigure[Bayes Magnitude\hspace{0.3in}] {\includegraphics[height=.18\textwidth]{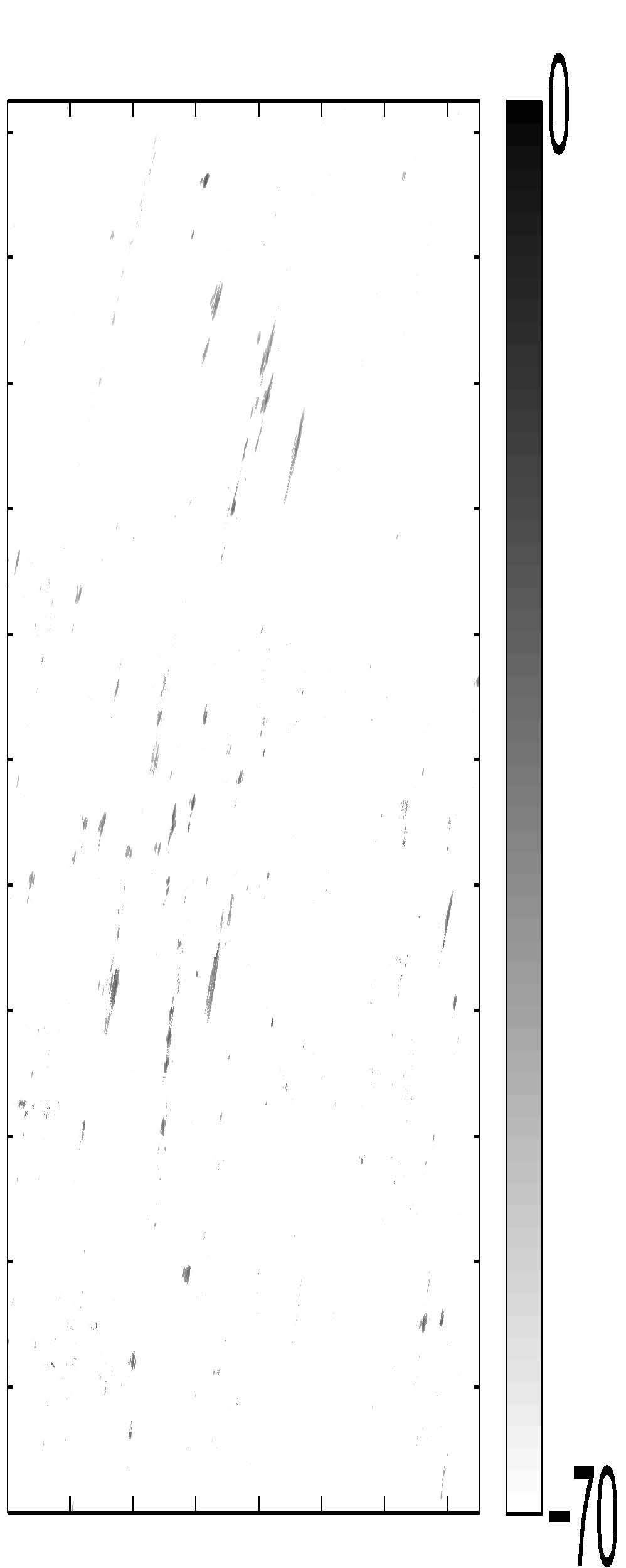}}
\subfigure[Bayes Phase\hspace{0.3in}] {\includegraphics[height=.18\textwidth]{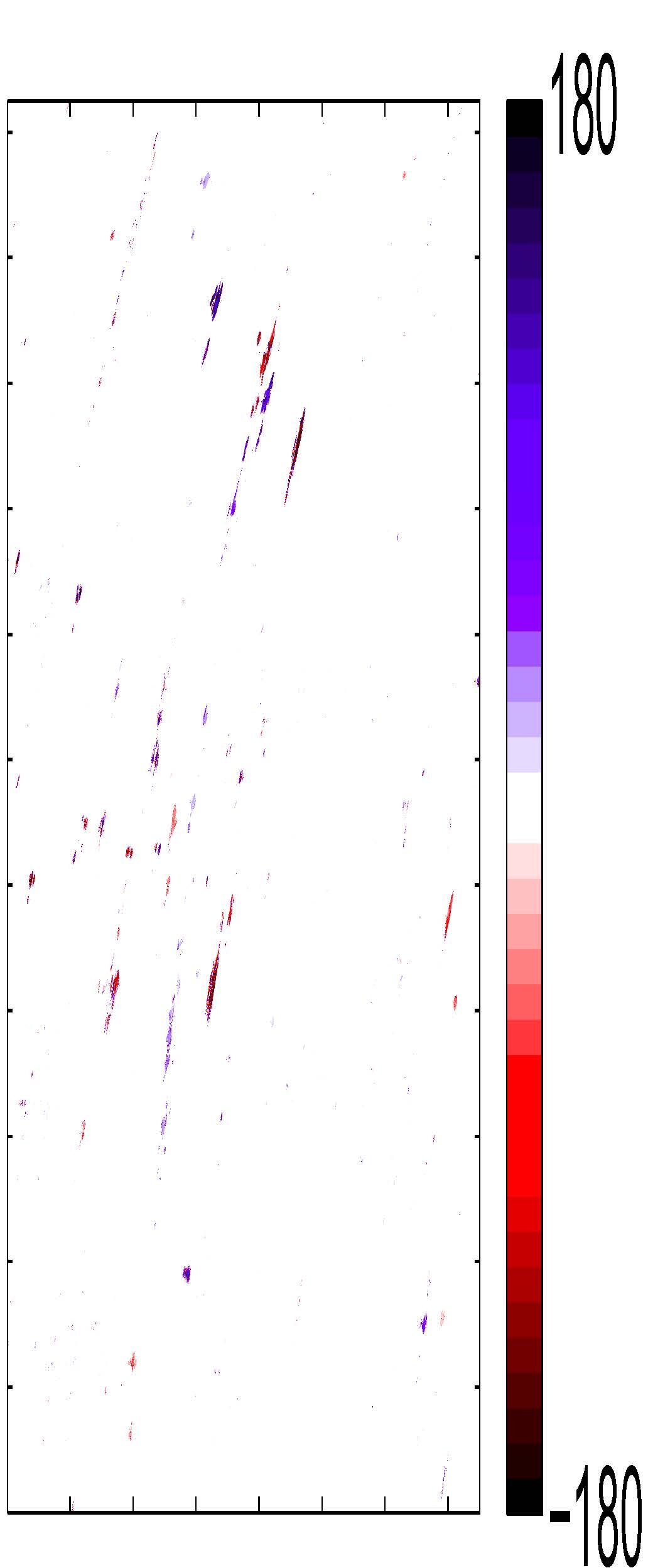}}
\subfigure[DPCA (original)\hspace{0.3in}] {\includegraphics[height=.18\textwidth]{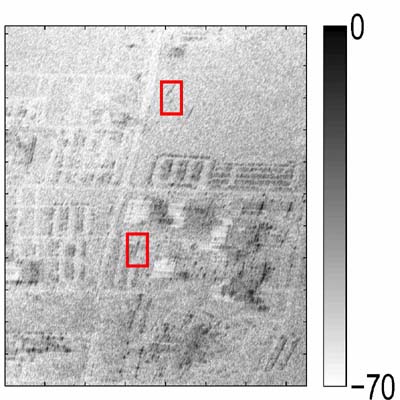}}
\subfigure[DPCA (calibrated)\hspace{0.3in}] {\includegraphics[height=.18\textwidth]{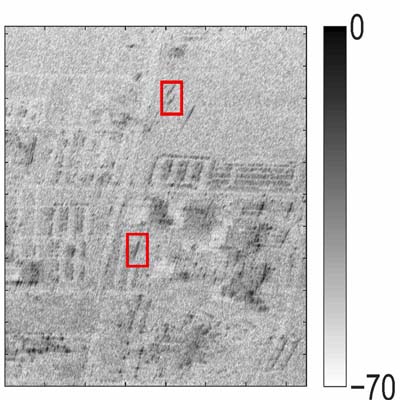}}
\subfigure[Phase (calibrated)\hspace{0.3in}] {\includegraphics[height=.18\textwidth]{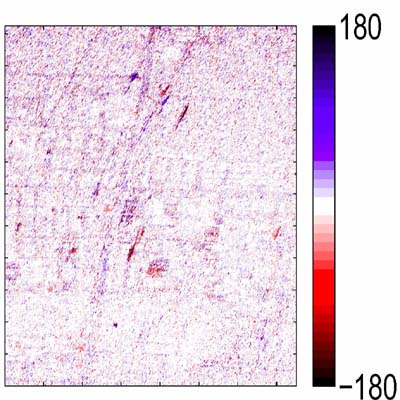}}
\caption{This figure compares the output of the proposed algorithm as a function of magnitude and phase for a scene of size 375m by 1200m and coherent processing interval of 0.5s.  The Bayes SAR algorithm takes the original SAR images in (a) and (b), estimates the nuisance parameters such as antenna miscalibrations and clutter covariances, and yields a sparse output for the target component in (c) and (d).   In contrast, the DPCA and ATI algorithms are very sensitive to the nuisance parameters, which make finding detection thresholds difficult.  In particular, consider the original interferometric phase image shown in (b).   It can be seen that without proper calibration between antennas, there is strong spatially-varying antenna gain pattern that makes cancellation of clutter difficult.   Calibration is generally not a trivial process, but to make fair comparisons to the DPCA/ATI algorithms, calibration in (f) and (g) is done by using the estimated coefficients $\sarHmat$ from the Bayes SAR algorithm.   In (e) and (f), the outputs of the DPCA algorithm are applied to the original images (all antennas) and the calibrated images (all antennas), respectively.  It should be noted that even with calibration, the DPCA outputs contain a huge number of false detections in high clutter regions.  Nevertheless, proper calibration enables detection of moving targets that are not easily detected without calibration, as highlighted by the red boxes.  Note that the Bayes SAR algorithm provides an output that is sparse, yet does not require tuning of thresholds as required by DPCA and/or ATI.} 
\label{sar-fig:comparisons-fullscene}
\end{figure*}

We begin by comparing the output of the proposed algorithm across the entire 375m by 1200m scene.  Figure \ref{sar-fig:comparisons-fullscene} shows the output of the Bayes SAR algorithm and the DPCA/ATI comparisons. It is seen that there are significant performance gains by using calibrated images as shown in (c) and (f) as compared to their original versions, (b) and (e), respectively.    Furthermore, the proposed approach also provides a sparse output without choosing thresholds as required by DPCA/ATI.  Note that in this figure, calibration is accomplished by using the outputs $\sarHmat$ from the Bayes SAR approach.

\detectionperffigswitch{
\begin{figure*}[thp]
\centering
\includegraphics[width=0.85\textwidth]{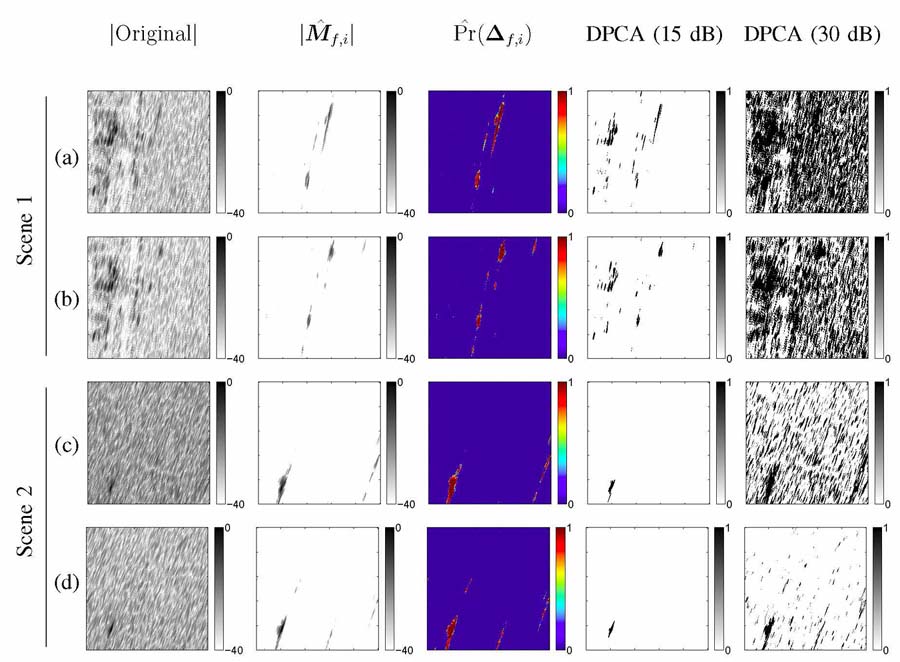}
\caption{This figure shows detection performance based on the magnitude of the target response with comparisons between the proposed Bayes SAR algorithm and displaced phase center array (DPCA) processing.  Note that DCPA declares a detection if the relative magnitude to the brightest pixel is greater than some threshold.  Results are given for two scenes of size 125m x 125m; within each scene, images were formed for two sequential 0.5 second intervals.   Scene 1 contains strong clutter in the upper left region, while Scene 2 has relatively little clutter. The columns of the figure provide from left-to-right: the magnitude of the original image, the estimated target component from the proposed algorithm, the probability of the target occupying a particular pixel, the output of DPCA with a relative threshold of 15 dB, and the output of DPCA with a relative threshold of 30 dB.  It is seen that DPCA has difficulty in canceling the clutter in Scene 1 with either threshold.  Moreover, in Scene 2 (c-d) DPCA misses detections of the low-magnitude target in the lower right for the 15 dB threshold.  In both scenes, there are many false alarms at the 30 dB threshold.  On the other hand, the proposed algorithm provides a sparse solution that detects all of these targets, while simultaneously providing a estimate of the probability of detection rather than an indicator output.}
\label{sar-fig:comparisons-magnitude}
\end{figure*}

\begin{figure*}[thp]
\centering
\input{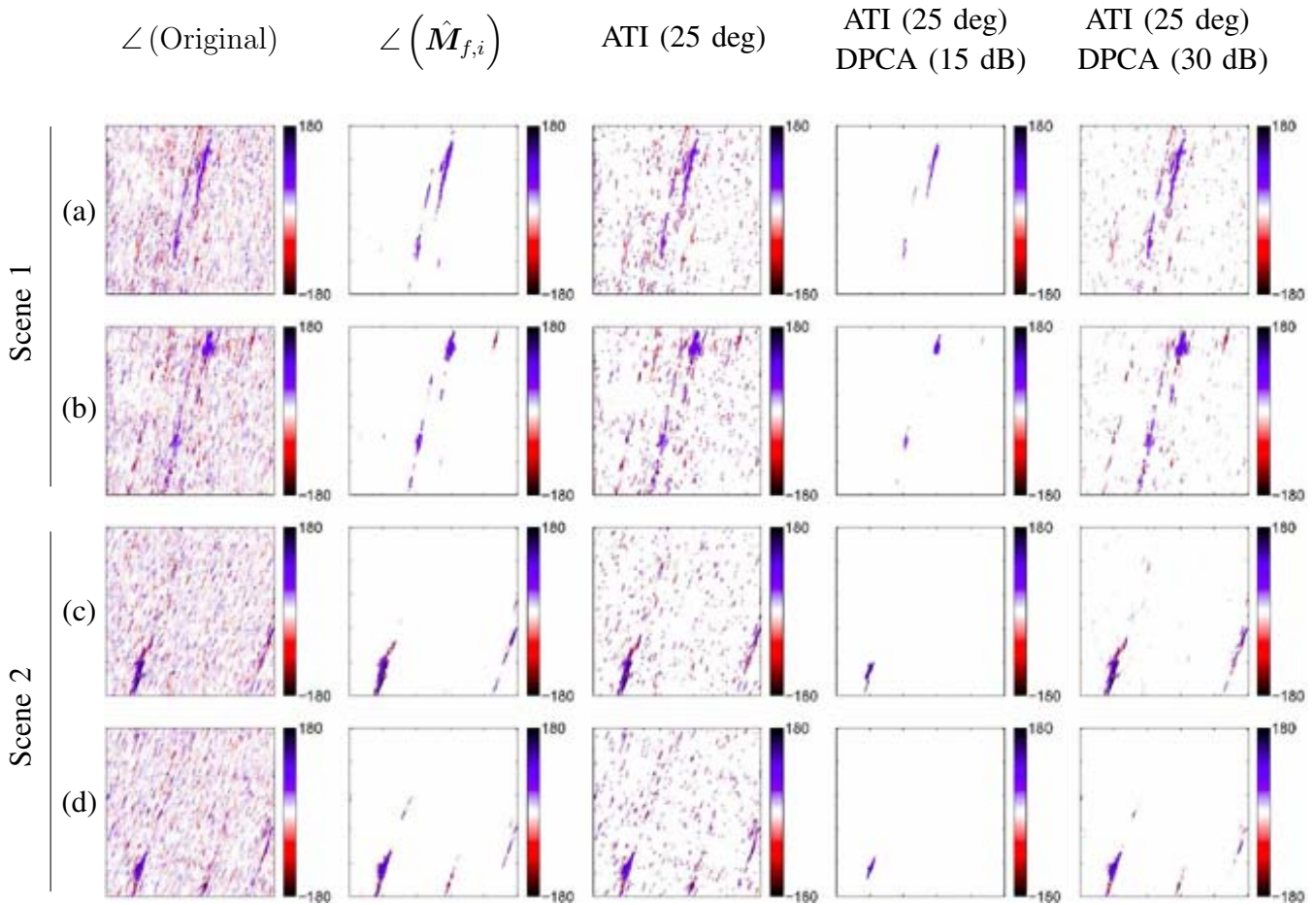}
\caption{This figure shows detection performance based on the phase of the target response with comparisons between the proposed algorithm, along-track interferometry (ATI) and a mixture algorithm between ATI/DPCA.  Results are given for the same two scenes in Figure \ref{sar-fig:comparisons-magnitude}.  In all cases, we show results for calibrated imagery where $\sarHmat$ are given by the output of the Bayes SAR algorithm, though this step is not trivial.  The columns of the figure provide from left-to-right: the phase of the image without thresholding, the estimated target phase component from the proposed algorithm, the output of ATI with a threshold of 25 degrees, the output of ATI/DPCA with (25 deg, 15 dB) thresholds, and the output of ATI/DPCA with (25 deg, 30 dB) thresholds.  In contrast to Figure \ref{sar-fig:comparisons-magnitude}, the contributions from the strong clutter are not very strong, though there are still numerous false alarms in the ATI and ATI/DPCA outputs.  It is seen that the ATI/DPCA combination with 15 dB magnitude threshold over-sparsifies the solution, missing targets in (b), (c), and (d).  On the other hand, the ATI/DPCA combination with 30 dB magnitude threshold detects these targets, but also includes false alarms in (a) and (b).
}
\label{sar-fig:comparisons-phase}
\end{figure*}

Figures \ref{sar-fig:comparisons-magnitude} and \ref{sar-fig:comparisons-phase} display the detection performance over two smaller scenes of size 125m by 125m as a function of magnitude and phase, respectively.  For each scene, images are provided for sequential scenes separated by 0.5 seconds.  Scene 1 contains strong clutter in the upper left region, while Scene 2 has relatively little clutter.  It is seen that the proposed approach (2nd column) provides a sparse solution containing the targets of interest in each of the 4 images.
Moreover, the 3rd column provides the estimated probability that a target occupies a given pixel, in comparison to the (0,1) output of DPCA.  Although most estimated probabilities are near 1, there are a few cases where this is not the situation: in scene 2(d), a low-magnitude target is detected with low probability in the lower-right; in scene 1(b) a few target pixels from the clutter region are detected with low probability.  In contrast, the performance of DPCA depends strongly on the threshold.  In Scene 1, a 30 dB threshold provides a large number of false alarms.  However, in Scene 2, the low-magnitude targets are missed for the 15 dB threshold, but detected at the 30 dB threshold.

Figure \ref{sar-fig:comparisons-phase} shows the detection performance based on phase over the same 4 images.  It is once again seen that the performance of the ATI/DPCA algorithms depend strongly on the thresholds, with performance that varies across thresholds from image to image.  On the other hand, the proposed approach is able to detect the targets with high fidelity regardless of the scene/image and does not require tuning of thresholds for detection.
}
{
\begin{figure*}[thp]
\centering
\input{comparisons-smallscene.tex}
\caption{This figure shows detection performance based on the magnitude/phase of the target response with comparisons between the proposed algorithm and displaced phase center array (DPCA) processing, and a mixture algorithm between DPCA and along-track interferometry (ATI).  Note that DCPA and ATI declare detections if the test statistic (magnitude for DPCA and phase for ATI) are than some threshold.  Results are given for two scenes of size 125m x 125m; within each scene, images were formed for two sequential 0.5 second intervals.   Scene 1 contains strong clutter in the upper left region, while Scene 2 has relatively little clutter.  The columns of the figure provide from left-to-right: the magnitude of the original image, the estimated probability of the target occupying a particular pixel (Bayes SAR), the estimated phase of the targets (Bayes SAR), the output of DPCA with a relative threshold of 30 dB, the output of ATI/DPCA with (25 deg, 15 dB) thresholds, and the output of ATI/DPCA with (25 deg, 30 dB) thresholds.  It is seen that without phase information to cancel clutter, DPCA (30 dB) contains an overwhelming number of false alarms for scenes (a-c), although the performance is reasonable for scene (d).  The ATI/DPCA algorithms provide sparser solutions by canceling the strong clutter.  It is seen that the ATI/DPCA combination with 15 dB magnitude threshold over-sparsifies the solution, missing targets in (b), (c), and (d).  On the other hand, the ATI/DPCA combination with 30 dB magnitude threshold detects these targets, but also includes numerous false alarms in (a) and (b).  On the other hand, the proposed algorithm provides a sparse solution that detects all of these targets, while simultaneously providing a estimate of the probability of detection rather than an indicator output.
}
\label{sar-fig:comparisons-smallscene}
\end{figure*}

Figure \ref{sar-fig:comparisons-smallscene} display the detection performance over two smaller scenes of size 125m by 125m as a function of magnitude and phase.  For each scene, images are provided for sequential scenes separated by 0.5 seconds.  Scene 1 contains strong clutter in the upper left region, while Scene 2 has relatively little clutter.  It is seen that the proposed approach (2nd and 3rd columns) provides a sparse solution containing the targets of interest in each of the 4 images.  Moreover, the 2nd column provides the estimated probability that a target occupies a given pixel, in comparison to the (0,1) output of DPCA and ATI.  Although most estimated probabilities are near 1, there are a few cases where this is not the situation: in scene 2(d), a low-magnitude target is detected with low probability in the lower-right; in scene 1(b) a few target pixels from the clutter region are detected with low probability.  In contrast, the performance of DPCA and ATI depend strongly on the threshold.  In (a-c), the DPCA-only output provides a large number of false alarms.  It is seen that the ATI/DPCA combination with 15 dB magnitude threshold over-sparsifies the solution, missing targets in (b), (c), and (d).  On the other hand, the ATI/DPCA combination with 30 dB magnitude threshold detects these targets, but also includes numerous false alarms in (a) and (b).  On the other hand, the proposed approach is able to detect the targets with high fidelity regardless of the scene/image and does not require tuning of thresholds for detection.
}

\subsubsection{Target motion models}
\begin{figure*}[thp]
\centering
\input{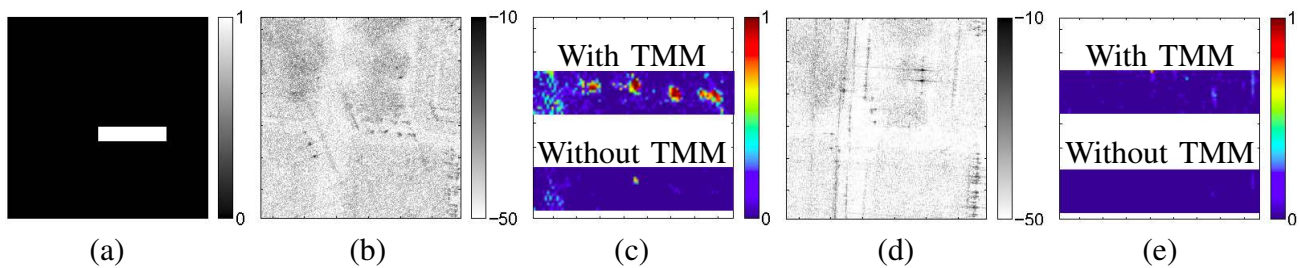}
\caption{This figure compares the performance of our proposed method with and without priors on target signature locations.  In this scene, targets are likely to be stopped at an intersection as shown by the region in (a).  A mission image containing targets is shown in (b) and a reference image without targets is shown in (d).  The estimated target probabilities are shown in (c) for the mission scene where inference was done both with/without a target motion model (TMM).  It can be seen that by including the prior information, we are able to detect stationary targets that cannot be detected from standard SAR moving target indication algorithms.  The estimated target probabilities in the reference scene are shown in (e), showing little performance differences when prior information is included in the inference.
}
\label{sar-fig:comparisons-priors}
\end{figure*}
Figure \ref{sar-fig:comparisons-priors} shows the output of the proposed approach when prior information on the location of the targets might
be available. For example, in the shown scene, targets are likely to be stopped at an intersection. The performance improvement is given for a mission scene that contains target in this high probability region.  On the other hand, there are no significant performance decreases in the reference scene that does not contain targets in the intersection region. This type of processing could be extended to a tracking environment, where targets are projected to likely be in a given location within the formed SAR image as discussed in Section \ref{sar-sec:markov-spatial-kinematic}.

\subsubsection{Estimation of radial velocity}
\journalTechSwitch{
\begin{figure}[th]
\centering
\includegraphics[height=.225\textwidth]{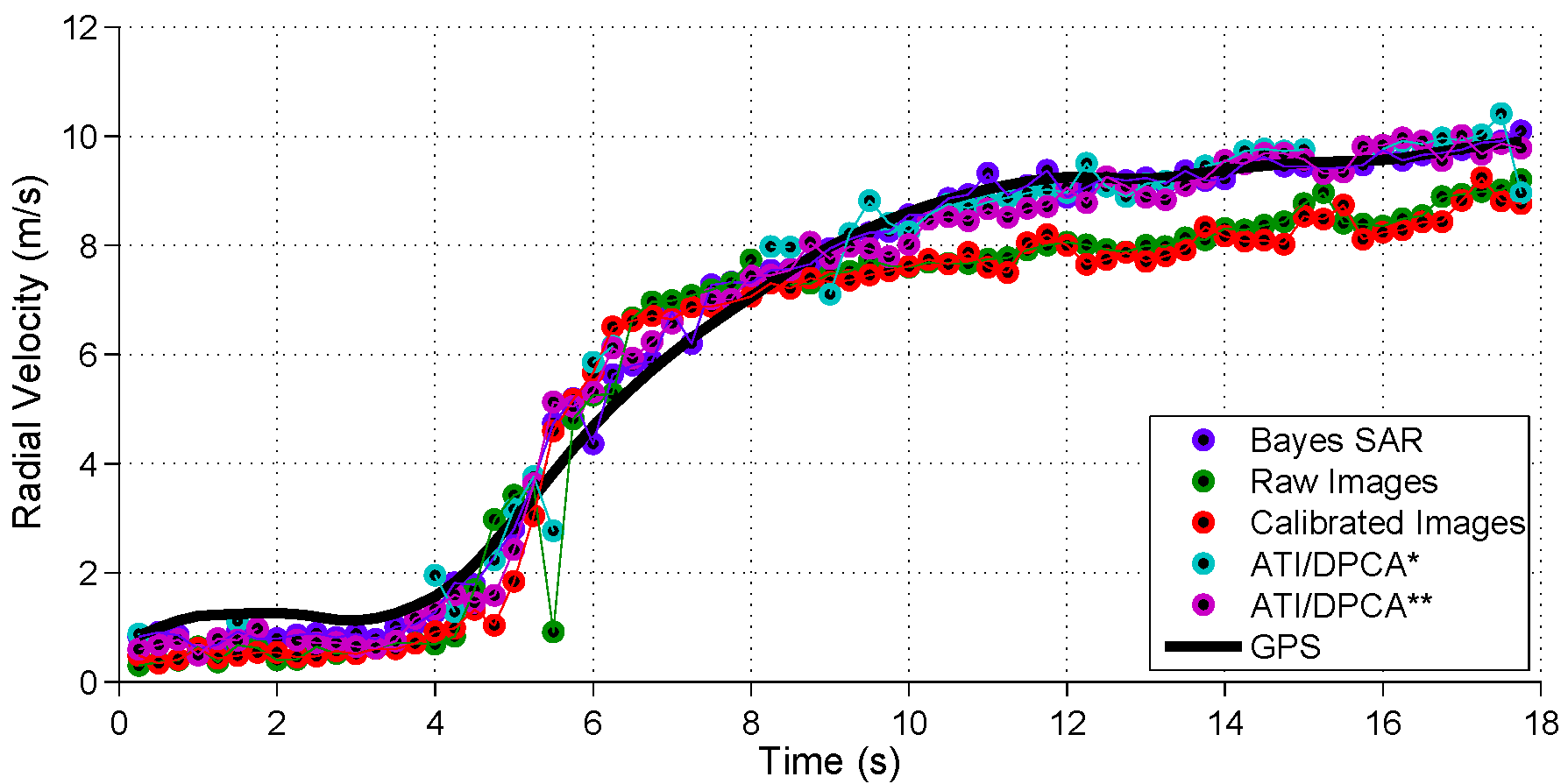}
\caption{This figure plots the estimated radial velocities (m/s) for a single target from measured SAR imagery over 18 seconds at 0.25 second increments.  Radial velocity, which is proportional to the interferometric phase of the pixels from multiple antennas in an along-track SAR system, is estimated by computing the average phase of pixels within a region specified by the GPS-given target state (position, velocity).  We compare the estimation of radial velocity from the output of the Bayes SAR algorithm, from the raw images, from the calibrated images (i.e, using the estimated calibration coefficients), and from two DPCA/ATI joint algorithms with phase/magnitude thresholds of (25 deg, 15 dB) and (25 deg, 30 dB) respectively.  For fair comparisons, the DPCA/ATI thresholds are applied to the calibrated imagery, though this is a non-trivial step in general.  The black line provides the GPS-truth.  \ignore{Numerical results are summarized in Table \ref{sar-table:radialvelcoity-estimation}.  It is seen that the Bayes SAR algorithm outperforms the others in terms of MSE for both targets.  Moreover, the Bayes SAR algorithm never misses a target detection in this dataset, which is not the case for the DPCA/ATI algorithms.}}
\label{sar-fig:radialvelocity-estimation}
\end{figure}}{
\begin{figure}[tp]
\centering
\subfigure[Target 1]{\includegraphics[height=.4\textwidth]{RadialVelocity_Targ2}}
\subfigure[Target 2]{\includegraphics[height=.4\textwidth]{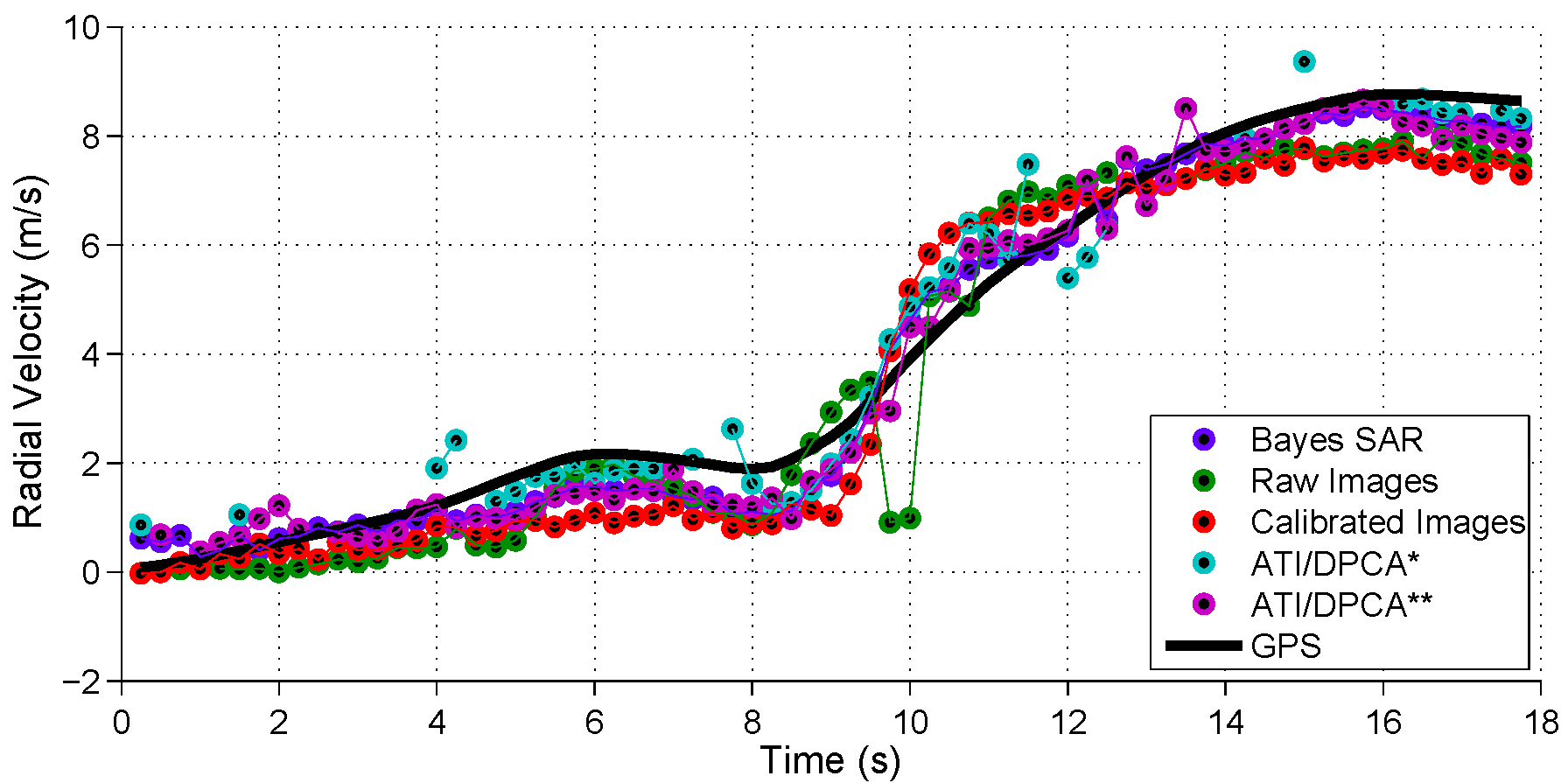}}
\caption{This figure plots the estimated radial velocities (m/s) for two targets from measured SAR imagery over 18 seconds at 0.25 second increments.  Radial velocity, which is proportional to the interferometric phase of the pixels from multiple antennas in an along-track SAR system, is estimated by computing the average phase of pixels within a region specified by the GPS-given target state (position, velocity).  We compare the estimation of radial velocity from the output of the Bayes SAR algorithm, from the raw images, from the calibrated images (i.e, using the estimated calibration coefficients), and from two DPCA/ATI joint algorithms with phase/magnitude thresholds of (25 deg, 15 dB) and (25 deg, 30 dB) respectively.  For best comparisons, the DPCA/ATI thresholds are applied to the calibrated imagery, though this is a non-trivial step in general.  The black line provides the GPS provided radial velocities.  Numerical results are summarized in Table \ref{sar-table:radialvelcoity-estimation}.  It is seen that the Bayes SAR algorithm outperforms the others in terms of MSE for both targets.  Moreover, the Bayes SAR algorithm never misses a target detection in this dataset, which is not the case for the DPCA/ATI algorithms.}
\label{sar-fig:radialvelocity-estimation}
\end{figure}
}

\renewcommand{\arraystretch}{1.25}
\journalTechSwitch{
\begin{table}[tp]
\caption{Radial velocity estimation (m/s) in measured SAR dataset.  The proposed algorithm (Bayes SAR) has lower bias and MSE, as well as fewer missed targets as compared to all other alternatives.  Moreover, all algorithms except `Raw' require additional calibrations between antennas, except the proposed algorithm which estimates calibration constants simultaneously with the target radial velocity.  Also, the proposed algorithm has nearly approximately half the error of the ATI/DPCA algorithms without requiring tuning of thresholds. }
\label{sar-table:radialvelcoity-estimation}
\centering
\begin{tabular}{|c|c|c|c|}
\hline
{Algorithm} &{Bias}  &{MSE}  &{No. Missed} \\
\hline\hline
{Raw}&0.56 &0.86 &7\\
\hline
{Calibrated}&0.60 &0.91 &0\\
\hline
{Bayes SAR}&0.11 &0.16 &0\\
\hline
{ATI/DPCA$^*$}&-0.06 &0.32 &57\\
\hline
{ATI/DPCA$^{**}$}&0.17 &0.24 &5\\
\hline
\end{tabular}
\end{table}
}
{
\begin{table}[tp]
\caption{Radial velocity estimation (m/s) in 2006 Gotcha collection dataset}
\label{sar-table:radialvelcoity-estimation}
\centering
\subtable[Target 1]{
\begin{tabular}{|c|c|c|c|}
\hline
{Algorithm} &{Bias}  &{MSE}  &{No. Missed} \\
\hline\hline
{Raw}&0.64 &0.94 &1\\
\hline
{Calibrated}&0.71 &1.02 &0\\
\hline
{Bayes SAR}&0.03 &0.10 &0\\
\hline
{ATI/DPCA$^*$}&-0.04 &0.20 &27\\
\hline
{ATI/DPCA$^{**}$}&0.10 &0.20 &2\\
\hline
\end{tabular}
}
\subtable[Target 2]{
\begin{tabular}{|c|c|c|c|}
\hline
{Algorithm} &{Bias}  &{MSE}  &{No. Missed} \\
\hline\hline
{Raw} & 0.47 &0.77 &6\\\hline
{Calibrated}&0.48 &0.79 &0\\\hline
{Bayes SAR}&0.19 &0.22 &0\\\hline
{ATI/DPCA$^*$}&-0.07 &0.43 &30\\\hline
{ATI/DPCA$^{**}$}&0.23 &0.28 &3\\\hline
\end{tabular}
}
\end{table}
}
\renewcommand{\arraystretch}{1.0}

The dataset used in this section contained a few GPS-truthed vehicles from which we can derive (a) the `true' location of the target within the formed SAR image, and (b) the target's radial velocity which is known to be proportional to the measured interferometric phase of the target pixels in an along-track system.
Figure \ref{sar-fig:radialvelocity-estimation} shows the estimated radial velocities for \journalTechSwitch{a single target}{two targets} over 18 seconds at 0.25 second increments.  We compare the estimation of radial velocity from the output of the Bayes SAR algorithm, from the raw images, from the calibrated images, and from two DPCA/ATI joint algorithms with phase/magnitude thresholds of (25 deg, 15 dB) and (25 deg, 30 dB) respectively.  For fair comparisons, the DPCA/ATI thresholds are applied to the calibrated imagery, though this is a non-trivial step in general.  Numerical results are summarized in Table \ref{sar-table:radialvelcoity-estimation}.  It is seen that the Bayes SAR algorithm outperforms the others in terms of MSE for both targets.  Moreover, the Bayes SAR algorithm never misses a target detection in this dataset, which is not the case for the DPCA/ATI algorithms.

\section{Discussion and future work}
\label{sar-sec:conclusion}
Recent work \cite{wright2009robust,lin2010augmented,candes2011robust} has shown that it is possible to successfully decompose natural high-dimensional signals/images into low-rank and sparse components in the presence of noise, leading to the so-called robust principal component analysis algorithms. \cite{ding2011bayesian} introduced a Bayesian formulation of the problem that built on the success of these algorithms with the additional benefits of (a) robustness to unknown densely distributed noise with noise statistics that can be inferred from the data, (b) convergence speeds in real applications of the mean solution that are similar to those of the optimization-based procedures, and (c) characterization of the uncertainty (i.e., estimates of the posterior distribution) that could lead to improvements in subsequent inference.  Moreover, the Bayesian formulation is shown to be capable of generalization to cases where additional information is available, e.g. spatial/Markov dependencies.
\ignore{
This chapter extends research in decomposing high-dimensional signals/images into low-rank and sparse components in the presence of noise \cite{wright2009robust,lin2010augmented,candes2011robust} to the case of separating target signatures from a low-dimensional clutter subspace in SAR imagery. In particular, we combine our understanding of the physical, kinematic, and statistical properties of SAR imagery into a single unified Bayesian structure that simultaneously (a) estimates the nuisance parameters such as clutter distributions and antenna miscalibrations and (b) extracts a sparse component containing the target signatures required for detection and estimation of the target state. Similar to Ding et al. \cite{ding2011bayesian}, this algorithm requires  few tuning parameters since most quantities of interest are inferred directly from the data - this allows the algorithm to be robust to a large collection of operating conditions. \ignore{ Moreover, we do not need to select thresholds for detection as in common algorithms such as DPCA and ATI, because we report detections as probabilities instead of (0, 1) outputs.}
The performance of the proposed approach is analyzed over both simulated and measured datasets, demonstrating competing or better performance than the robust PCA algorithms and ATI/DPCA.
\perfpredswitch{}{Moreover, it is shown that the outputs of the Bayesian inference can be used for future performance prediction through examples of derived likelihood ratio tests and Cram\' er-Rao Lower Bounds for spatial errors}.
}
Future work will include the development of algorithms that exploit the use of a posterior distribution for improved performance in a signal processing task, e.g. detection, tracking or classification.  In particular, we are interested in using algorithms for simultaneously detecting and estimating targets over a sparse scene with resource constraints \ignore{as discussed in Chapters \ref{chap:arap} and \ref{chap:darap}}, as well determining the fundamental performance limits of a SAR target tracking system.  Furthermore, we would also like to consider other generalizations to the SAR image model, such as complex target maneuvers, multiple target classes, and explicit tracking of the target phase.

\ignoreForJournal{
\appendices

\section{Target signature prediction}
\label{sar-appendix:signature-prediction}
In some applications, such as target tracking or sequential detection, we may have access to an estimate of the kinematic state of the target(s) of interest, such as position, velocity and acceleration.  This may be useful for predicting the location of the target at sequential frames.  For simplicity, consider a single target whose state $(\vr(\tau),\vdotr(\tau))$ is known with standard errors $(\sigma_r,\sigma_{\dot{r}})$, where $\tau$ denotes the slow-time (i.e., time of the radar pulse).  In standard SAR image formation, moving targets tend to appear displaced and defocused in as described in the literature by Fienup \cite{fienup2001detecting} and Jao \cite{jao2001theory}.  Moreover, Jao shows that given the radar trajectory $(\vq,\vdotq)$ and the target trajectory $(\vr,\vdotr)$, one can predict the location of the target signature within the image ${\bm p}$ by solving a system of equations that equate Doppler shifts and ranges, respectively, at each pulse:
\begin{align}
\frac{d}{d\tau}\left[\norm{\vp-\vq(\tau)}_2 - \norm{\vr(\tau)-\vq(\tau)}_2\right]_{\vp=\vp^*}=0\\
\norm{\vp^*-\vq(\tau)}_2=\norm{\vr(\tau)-\vq(\tau)}_2,
\end{align}
which can be reduced to the simpler system of equations:
\begin{align}
\vdotq(\tau)\cdot[\vp^*-\vq(\tau)] &= [\vdotr(\tau)-\vdotq(\tau)]\cdot[\vr(\tau)-\vq(\tau)]\\
\norm{\vp^*-\vq(\tau)}_2&=\norm{\vr(\tau)-\vq(\tau)}_2
\end{align}
In practice, the target state $(\vr,\vdotr)$ is unknown or known with some uncertainty.  In the latter case, we can predict the probable locations of the target signature by one of several methods, including:
\begin{itemize}
\item Monte Carlo estimation of the target signature locations.
\item Gaussian approximation using linearization or the unscented transformation.
\item Analytical approximation as proposed by Newstadt et al. \cite{News2010-UA-1}.
\end{itemize}
\subsection{Notation}
Following the derivation of Jao \cite{jao2001theory}, we will assume the following notation:
\begin{itemize}
\item ${\bf r}(\tau) = (r_x,r_y,r_z)$ is the position of a point scatterer.
\item ${\bf q}(\tau) = (q_x,q_y,q_z)$ is the position of the radar platform.
\item $\dot{\bf r}(\tau) = (\dot{r}_x,\dot{r}_y,\dot{r}_z)$ is the velocity of a point scatterer.
\item $\dot{\bf q}(\tau) = (\dot{q}_x,\dot{q}_y,\dot{q}_z)$ is the true position of the platform.
\item ${\bf p} = (p_x,p_y,p_z)$ is a pixel location within the image.
\item $\tau$ represents the slow-time (i.e., pulse of the radar sample).
\end{itemize}
\subsection{Deterministic solution}
In the deterministic case, where ${\bf r},\dot{\bf r},{\bf q}$, and $\dot{\bf q}$ are all known, we can find the pixel ${\bm p}^*$ where the target signature will be focused at time $\tau$ by solving equations (\ref{sar-eq:equal-doppler}) and (\ref{sar-eq:equal-range}).  In particular, if we assume that $z$-coordinate is given by a function
\begin{equation}
p_z = h(p_x,p_y),
\end{equation}
then we can give explicit expressions for $(p_x,p_y,p_z)$ in some cases of $h$.  We will focus on the simple case where $h(p_x,p_y) = z_0$ (i.e, constant elevation), though this can be easily extended to other cases (for example, with a depth elevation map).

To solve the system of equations, let
\begin{align}
\alpha(\tau) &= \norm{{\bf r}(\tau) - {\bf q}(\tau)}_2^2\\
\beta(\tau) &= \dot{\bf q}(\tau)\cdot{\bf r}(\tau) - \dot{\bf r}(\tau)\cdot({\bf r}(\tau)-{\bf q}(\tau))
\end{align}
Then we have
\begin{equation}
\label{sar-tech-eq:pred-alpha}
\begin{split}
\alpha(\tau) &= \norm{{\bf p}^*-{\bf q}(\tau)}_2^2\\
&= \left(p_x^*-q_x(\tau)\right)^2 + \left(p_y^*-q_y(\tau)\right)^2 + \left(z_0-q_z(\tau)\right)^2
\end{split}
\end{equation}
and re-arranging equation (\ref{sar-eq:equal-doppler}) we have
\begin{equation}
\begin{split}
\beta(\tau) &= \dot{\bf q}(\tau)\cdot{\bf p}^*\\
&= \dot{q}_x p_x^* + \dot{q}_y p_y^* + \dot{q}_z z_0
\end{split}
\end{equation}
For this derivation, assume that $\dot{q}_x\neq 0$\footnote{By assumption, the radar has non-zero velocity in the $xy$-plane.  Thus, if $\dot{q}_x=0$, then this derivation should be valid if we switch the $x$ and $y$ indices.}.  Therefore, solving for $p_x^*$, we get:
\begin{equation}
\label{sar-tech-eq:pred-px}
\begin{split}
p_x^* &= \frac{\beta(\tau)-\dot{q}_y p_y^* - \dot{q}_z z_0}{\dot{q}_x}\\
&= \left(\frac{\beta(\tau) - \dot{q}_z z_0}{\dot{q}_x}\right) + \left(-\frac{\dot{q}_y}{\dot{q}_x}\right)p_y^*\\
&= \gamma_0 + \gamma_1 p_y^*
\end{split}
\end{equation}
Plugging into equation (\ref{sar-tech-eq:pred-alpha}) we get:
\begin{equation}
\begin{split}
\alpha(\tau) = \left(\gamma_0 + \gamma_1 p_y^* - q_x(\tau)\right)^2 + \left(p_y^*-q_y(\tau)\right)^2 + \left(z_0-q_z(\tau)\right)^2
\end{split}
\end{equation}
which can be re-arranged as
\begin{equation}
\begin{split}
a_y(p_y^*)^2 + b_yp_y^* + c_y(p_y^*)^2 = 0
\end{split}
\end{equation}
where
\begin{equation}
\begin{split}
a_y &=\left(1+\gamma_1^2\right)\\
b_y &= 2\left(\gamma_0\gamma_1-\gamma_1q_x(\tau)-q_y(\tau)\right)\\
c_y &= \left(\gamma_0-q_x(\tau)\right)^2 + (q_y(\tau))^2  + \left(z_0-q_z(\tau)\right)^2 - \alpha(\tau)\\
& = \left(\gamma_0^2 + z_0^2\right) - 2\left(\gamma_0q_x(\tau)+z_0q_z(\tau)\right) + \norm{{\bf q}(\tau)}_2^2 -  \norm{{\bf r}(\tau) - {\bf q}(\tau)}_2^2
\end{split}
\end{equation}
Then $p_y^*$ is given by the solution of the quadratic equation:
\begin{equation}
\label{sar-tech-eq:pred-py}
p_y^* = \frac{-b_y \pm \sqrt{b_y^2-4a_yc_y}}{2a_y}
\end{equation}
and $p_x^*$ is given by equation (\ref{sar-tech-eq:pred-px}).  This solution suggests that the target energy will generally actually appear at two locations.  However, in most cases only one of these locations will be in the formed SAR image.  Thus, we generally choose the solution $(p_x^*,p_y^*)$ that is closest to the scene center $(0,0)$.

Finally, we note that equations (\ref{sar-tech-eq:pred-px}) and (\ref{sar-tech-eq:pred-py}) provide the pixel location containing the target energy at a single pulse time, $\tau$.  Generally, images are formed by integrating pulses over a coherent processing interval (CPI) containing multiple times $\tau\in[T_0,T_1]$. 

\subsection{Uncertainty model}
It is unlikely that we will have perfect information regarding the target state ${\bm r}(\tau)$ and $\dot{\bm r}(\tau)$ at all times $\tau$.  On the other hand, there are special cases where we might have some information about these states that include
\begin{itemize}
\item A tracking environment, where we estimate target position and velocities with standard errors, $\sigma_r$ and $\sigma_{\dot{r}}$, respectively.
\item A persistently monitored scene, where we have knowledge of traffic patterns or road systems.  In such a case, we have prior knowledge of likely target states.
\end{itemize}
In either case, we have a characterization of likely behavior of the target kinematic state.  We can represent this knowledge in many ways that could include
\begin{itemize}
\item A linear kinematic model, where
\begin{equation}
\begin{split}
{\bm r}(\tau) &= {\bm r}_0 + {\bm v}\tau + {\bm a}\tau^2/2\\
 {\bm r}_0 &\sim N\left({\bm \mu}_r, \sigma_r^2{\bm I}\right)\\
 {\bm v} &\sim N\left({\bm \mu}_v, \sigma_v^2{\bm I}\right)\\
 {\bm a} &\sim N\left({\bm \mu}_a, \sigma_a^2{\bm I}\right)
\end{split}
\end{equation}
\item A random kinematic model, where at each time $\tau$
\begin{equation}
\begin{split}
{\bm r}(\tau) &\sim N\left({\bm \mu}_r(\tau), [\sigma_r(\tau)]^2{\bm I}\right)\\
\dot{\bm r}(\tau) &\sim N\left({\bm \mu}_{\dot{r}}(\tau), [\sigma_{\dot{r}}(\tau)]^2{\bm I}\right)\\
\end{split}
\end{equation}
\end{itemize}
Note that both models assume that the position and velocity vectors are Gaussian distributed.  However, the first model is characterized by only 6 random variables (2 each for position, velocity, and acceleration) regardless of the number of pulses.  The second model, on the other hand, assumes 4 random variables for each pulse $\tau$.  In fact, the first model can be seen as a specialization of the second model for specific structures for the mean and variance parameters.

In this document, the choice of target kinematic model depends on the inference method which we will use to derive the distribution of the target locations.  In Monte Carlo sampling, the choice of model is of relatively insignificant computational burden as compared to the generation of the Monte Carlo samples.  On the other hand, in the analytical approximation methods, the choice of target kinematic model is of great importance.

The goal of this section is to provide a prediction model for the locations of targets within a SAR image given a target kinematic model.  In particular, we define this model through the distribution of pixel locations:
\begin{equation}
f(p_x,p_y)
\end{equation}
which is assumed to have support on $\mathbb{R}^2$.  Generally, images are formed on a discrete grid so that we should really consider a discrete distribution.  However, for simplicity we consider a continuous domain in this section.  

Moreover, we have to be careful how we define the distribution of pixel locations for SAR images formed by integrating multiple radar pulses.  Consider a probability distribution function (PDF) for the target location at pulse $\tau_i$ given by $f(p_x(\tau_i),p_y(\tau_i))$.  We define the target distribution of interest as:
\begin{equation}
\label{sar-tech-eq:f_px_py}
f(p_x,p_y)  = \frac{1}{T}\sum_{i=1}^T f(p_x(\tau_i),p_y(\tau_i))
\end{equation}
This is equivalent to the distribution of the target occupying location $(p_x,p_y)$ at any of $T$ integrated pulses.  We could consider a richer description by solving for the joint distribution on $\set{p_x(\tau_i),p_y(\tau_i)}_{i=1}^T$.  However, this will be generally very high-dimensional and might not provide any additional benefit over the distribution given by equation (\ref{sar-tech-eq:f_px_py}) be useful for the purposes described in the paper.

Finally, since we will solve for the distributions $f(p_x(\tau_i),p_y(\tau_i))$ independently for each $\tau_i$, we will consider the random kinematic model only.  In the naive situation where the distributions of target locations are independent over time, this will provide solutions that can be approximated analytically with just a few minor assumptions.

\subsection{Monte Carlo prediction}
The most straightforward way to approximate the distribution in equation (\ref{sar-tech-eq:f_px_py}) is to use Monte Carlo sampling from the linear/random target kinematic models, followed by projection of those target states into the image domain using equations (\ref{sar-tech-eq:pred-px}) and (\ref{sar-tech-eq:pred-py}).  A Monte Carlo representation is subsequently given by the average number of samples occupying any pixel.  Note that since the Monte Carlo representation will contain discrete samples, we will end up with a discrete probability mass function (PMF) rather than a PDF.  

\subsection{Gaussian approximation}
\label{sar-tech-subsec:gaussian}
Rather than using a potentially high-dimensional PMF or PDF representation of the target location at time $\tau_i$, we could consider a Gaussian approximation that represents the probability distribution with just two parameters: the mean ${\bm \mu}_i\in\mathbb{R}^2$ and the covariance ${\bm \Sigma}_i\in\mathbb{R}^{2\times 2}$ for each slow-time.  Then the PDF is given by a Gaussian mixture model of form:
\begin{equation}
\label{sar-tech-eq:f_px_py_gmm}
f({\bm p}) \approx \frac{1}{T}\sum_{i=1}^T \normpdf \left({\bm p};{\bm \mu}_i, {\bm \Sigma}_i\right)
\end{equation}
where $\normpdf\left({\bm x};{\bm \mu}, {\bm \Sigma}\right)$ is the multivariate normal distribution PDF of ${\bm x}$ with mean ${\bm \mu}$ and covariance ${\bm \Sigma}$.

To find the means ${\bm \mu}_i$ and covariances ${\bm \Sigma}_i$, we could consider linearization of the solution to equations (\ref{sar-tech-eq:pred-px}) and (\ref{sar-tech-eq:pred-py}) around a particular state ${\bm \xi}(\tau_i)\trieq\set{{\bm r}(\tau_i),\dot{\bm r}(\tau_i)}$.  This is akin to the approximation made by the Extended Kalman Filter, where the state is Gaussian but the observations are non-linear.  Let us define 
\begin{equation}
g({\bm \xi}(\tau_i)) = \left[{\begin{array}{*{20}c}
  {p_x^*({\bm \xi}(\tau_i))}  \\
   {p_y^*({\bm \xi}(\tau_i))}   \\
\end{array}}\right]
\end{equation}
where $p_x^*$ and $p_y^*$ are given by equations (\ref{sar-tech-eq:pred-px}) and (\ref{sar-tech-eq:pred-py}).  Then the first-order linearization around the mean ${\bm \mu_\xi}$ is given by:
\begin{equation}
G({\bm \xi}) = g({\bm \mu_\xi}) + \left(\nabla g({\bm \xi})\big|_{{\bm \xi}={{\bm \mu_\xi}}}\right)^T({\bm \xi}-{\bm \mu_\xi})
\end{equation}
Since this is a linear function of a Gaussian distributed vector in ${\bm \xi}(\tau_i)$, we know that the pixels ${\bm p}(\tau_i)\approx G({\bm \xi}(\tau_i))$ are distributed as
\begin{equation}
{\bm p}(\tau_i) \sim N\left( g({\bm \mu_\xi(\tau_i)}), \left(\nabla g({\bm \xi})\big|_{{\bm \xi}={{\bm \mu_\xi}(\tau_i)}}\right)^T {\bm \Sigma}_{\bm \xi}(\tau_i) \left(\nabla g({\bm \xi})\big|_{{\bm \xi}={{\bm \mu_\xi}(\tau_i)}}\right)     \right) ,
\end{equation}
where ${\bm \Sigma}_{\bm \xi}(\tau_i)$ is given by
\begin{equation}
{\bm \Sigma}_{\bm \xi}(\tau_i) =  \left[{\begin{array}{*{20}c}
  {{\bm \Sigma}_r(\tau_i)} & {0}  \\
   {0} & {{\bm \Sigma}_{\dot{r}}(\tau_i)}    \\
\end{array}}\right]
\end{equation}

\subsection{Analytical approximation}
It is also possible to get a closer approximation than the linearization example provided above by doing some analytical derivations.  In particular, let us assume that the radar platform moves in the $x$-direction so that $\dot{q}_y=\dot{q}_z=0$ and $|\dot{q}_x|>0$.  In the general case, this derivation would hold for a transformed set of coordinates $(p_x',p_y')$, though we won't go into that derivation here.  In the former case, equations 
(\ref{sar-tech-eq:pred-px}) and (\ref{sar-tech-eq:pred-py}) reduce to:
\begin{align}
\label{sar-tech-eq:pred-px-simple}
p_x^* &= \frac{\beta(\tau)}{\dot{q}_x}=\frac{\dot{q}_x r_x - (r_x-q_x)v_x - (r_y-q_y)v_y}{\dot{q}_x} \\
\label{sar-tech-eq:pred-py-simple}
p_y^* &= q_y \pm \sqrt{\alpha(\tau)-(p_x-q_x)^2-(z_0-q_z)^2}
\end{align}
From equation (\ref{sar-tech-eq:pred-px-simple}), we see that
\begin{equation}
f(p_x^*|r_x,r_y) \sim N(\mu,\sigma^2)
\end{equation}
where
\begin{align}
\label{sar-tech-eq:pred-mu-rxry}
\mu &= \frac{1}{\dot{q}_x}\left([\dot{q}_x-\mu_{vx}]r_x - \mu_{vy}r_y + q_x\mu_{vx}+q_y\mu_{vy}\right)\\
\sigma^2 &= \frac{(r_x-q_x)^2 + (r_y-q_y)^2}{(\dot{q}_x)^2}\sigma_v^2
\end{align}
where we have assumed that $\Sigma_{\dot{r}} = \sigma_v^2{\bm I}$.  Note that $\sigma^2$ is a function of the position $\bm r$.  However, since $\norm{q}\gg\norm{r}$ in general, we make a zero-th order approximation here so that
\begin{equation}
\sigma^2 \approx \sigma_0^2 \trieq \frac{(\mu_{rx}-q_x)^2 + (\mu_{ry}-q_y)^2}{(\dot{q}_x)^2}\sigma_v^2
\end{equation}
In this case, we see that we can find
\begin{equation}
\begin{split}
f(p_x) &= \int_{-\infty}^\infty \int_{-\infty}^\infty f(p_x|r_x,r_y) f(r_x,r_y) dr_xdr_y\\
& = \intinf\intinf \normpdf(p_x; \mu(r_x,r_y), \sigma_0^2)\normpdf(r_x; \mu_{rx}, \sigma_r^2)  \normpdf(r_y; \mu_{ry}, \sigma_r^2) dr_x dr_y
\end{split}
\end{equation}
where we have assumed that $\Sigma_{{r}} = \sigma_r^2{\bm I}$.  Since $\mu(r_x,r_y)$ is linear in $r_x$ and $r_y$ from equation (\ref{sar-tech-eq:pred-mu-rxry}), we can analytically solve this integral to see that 
\begin{equation}
p_x^* \sim N(\mu_{px}, \sigma_{px}^2)
\end{equation}
where
\begin{align}
\mu_{px} &=  \frac{1}{\dot{q}_x}\left(q_x\mu_{vx} + q_y\mu_{vy} + \dot{q}_x\mu_{rx}-\mu_{vx}\mu_{rx}-\mu_{vy}\mu_y\right)\\
\sigma_{px}^2 &=  \frac{\sigma_r^2}{(\dot{q}_x)^2}\left[\left(\dot{q}_x-\mu_{vx}\right)^2+\mu_{vy}^2\right]+\sigma_0^2
\end{align}
Since $p_y^*$ in equation (\ref{sar-tech-eq:pred-py-simple}) is non-linear, we make one more assumption with a first-order linearization around $p_y^*({\bm \mu_r})$. Note note that given $p_x^*$, equation (\ref{sar-tech-eq:pred-py-simple}) only depends on the target state through $\bm r$ (and not on the velocity $\dot{\bm r}$).  Define $s({\bm r},p_x)$ to be the value of equation (\ref{sar-tech-eq:pred-py-simple}) given state $\bm r$ and pixel location $p_x$.  Then we approximate $p_y^*$ as
\begin{equation}
p_y^*(\bm r) | p_x^* \approx s({\bm \mu_r},p_x^*) + \left(\nabla  s({\bm \mu_r},p_x^*)\big|_{{\bm r}={{\bm \mu_r}}}\right)^T({\bm r}-{\bm \mu_r})
\end{equation}
Finally, we note that given $p_x$, $p_y$ is Gaussian distributed with distribution:
\begin{align}
p_y|p_x &\sim N(\mu_{py},\sigma_{py}^2)\\
\mu_{py} &=   s({\bm \mu_r},p_x^*) \\
\sigma_{py}^2 &= \sigma_r^2 \left(\nabla s({\bm \mu_r},p_x^*) \big|_{{\bm r}={{\bm \mu_r}}}\right)^T \left(\nabla  s({\bm \mu_r},p_x^*) \big|_{{\bm r}={{\bm \mu_r}}}\right)
\end{align}

Note that both $p_x^*$ and $p_y^*\ |\ p_x^*$ have Gaussian distributions that can be described by a mean and covariance term.  In contrast to Section \ref{sar-tech-subsec:gaussian}, the distribution is not jointly Gaussian because the mean and covariance of $p_y^*\ |\ p_x^*$ depend on $p_x^*$.  Nevertheless, one can easily evaluate this PDF at any pixel $(p_x,p_y)$.  Over short CPIs, both approximations will probably lead to similar results.

\section{Inference Details}
\label{sar-appendix:inference}
In the hierarchical model proposed in Section \ref{sar-sec:image-model}, the distribution of hyper-parameters at the base layer are generally chosen to be conjugate to the distributions at the next layer.  This allows for efficient approximation methods for the posterior distribution in the sense that we can sample exactly from these distributions.  In particular, we use a Markov Chain Monte Carlo (MCMC) algorithm in the form of a Gibbs sampler to iteratively estimate the full joint posterior.  In MCMC, this distribution is approximated by drawing samples iteratively from the conditional distribution of each (random) model variable given the most recent estimate of the rest of the variables (which we denote by $-$).  Let ${\bm \Theta}=\set{{\bm B},{\bm X},{\bm G},{\bm M},{\bm \Delta}^G,{\bm \Delta}^M,{\bm H},{\bm C},{\bm \eta}}$ represent a current estimate of all of the model variables where ${\bm \eta}$ represents the set of all hyper-parameters.  Given measurements $\bm I$, the inference algorithm is given in Figure \ref{sar-alg:gibbs-sampler}.  Note that MCMC algorithms require a burn-in period after the Markov chain has become stable, where the duration of burn-in period depends on the problem. After this point, we collect $N_{samples}$ samples that represent the full joint distribution.

The sampling details are provided for each of the steps in Figure \ref{sar-alg:gibbs-sampler} individually.

\subsection{Basic Decomposition}
Given the parameters in Tables \ref{sar-table:hyperparameters-distributional-models-covariances} and \ref{sar-table:hyperparameters-distributional-models-other}, we arrive at one of the primary benefits of using Gibbs Sampling for inference: namely that we can independently sample across pixels and frames.  In distributional form, we have
\begin{align}
\label{sar-app-eq:samplestep-basic}&f({\bm B},{\bm X},{\bm G},{\bm M},{\bm \Delta}^G,{\bm \Delta}^M|{\bm I,-})\\
\nonumber&= \prod\limits_{p,f} f(\sarBv,\sarXvi,\sarGvi,\sarMvi,\sarDG,\sarDMi|{\bm I},-)
\end{align}
The conditional independence among pixels and frames given the nuisance parameters allows us to easily parallelize the sampling procedure over the largest dimensions of the state.  Moreover, we can extend the parallelization to sampling independently over passes by separating the sampling of equation (\ref{sar-app-eq:samplestep-basic}) into two Gibbs steps from the densities:
\begin{align}
\label{sar-app-eq:samplestep-basic2}
&f(\sarBv,\sarXvi,\sarMvi,\sarDMi|{\bm I},-)\\
\nonumber&\qquad= f(\sarBv|{\bm I},\sarXvi,\sarMvi,\sarDMi,-)\\
\nonumber&\qquad\cdot\prod_{i} f(\sarXv,\sarMv|\sarDM,{\bm I},-)f(\sarDM|{\bm I},-)\\
\label{sar-app-eq:samplestep-basic3}
&f(\sarGvi,\sarDG|{\bm I},-)\\
\nonumber&\qquad=f(\sarDG|{\bm I},-)\prod_{i} f(\sarGv|{\bm I},\sarDG,-)
\end{align}
It should be noted that each of these distributions have explicit forms as either multivariate Gaussian or Bernoulli distributed.  For example consider the distribution
\begin{equation}
f(\sarXv,\sarMv|\sarDM,{\bm I},-) = f(\sarXv,\sarMv|\sarDM,\sarRv,-)
\end{equation}
where we define 
\begin{equation}
\sarRv \trieq \left(\sarIv ./ \sarHv\right) - \sarDG \sarGv,
\end{equation} 
where $./$ is the point-wise division operator. In particular, we know that
\begin{equation}
\sarRv = \sarBv + \sarXv + \sarDM \sarMv + \sarVv,
\end{equation}
Thus, given $\sarDM$, we have $\sarRv$ as a linear combination of Gaussian random variables.  Thus, by standard conditional distributions of a Gaussian random vector, we know that the distribution of $\sarBv$ and $\sarXv$ must also be Gaussian.  Table \ref{sar-table:gaussian-parameters-inference} gives the means and covariances of the Gaussian distributions in equations (\ref{sar-app-eq:samplestep-basic2}) and (\ref{sar-app-eq:samplestep-basic3}).  Moreover, the distributions of the indicator functions are easily found by Bayes rule.  For example, define the quantity
\begin{equation}
z_d = f(\sarIv|\sarDM=d,\sarHv,\sarDG,\sarGv,\sarBv).
\end{equation}
Then it is easily seen that for the target indicators we have by Bayes rule
\begin{equation}
\begin{split}
f(\sarDM=1|{\bm I},-) &= \frac{z_1 f(\sarDM=1)} {\sum\limits_{d=(0,1)}z_d f(\sarDM=d)}\\
&= \left[1 + \frac{1-\pi_{f,i}^M(p)}{\pi_{f,i}^M(p)}\frac{z_0}{z_1}   \right]^{-1}
\end{split}
\end{equation}
Note that $z_d$ is just an evaluation of the Normal PDF so that it can be simply calculated.  Table \ref{sar-table:bernoulli-parameters-inference} provides explicit values of $z_d$ for both the target and glint indicators, where $\normpdf(x;{\bm \mu},{\bm \Sigma})$ is the multivariate normal PDF with mean ${\bm \mu}$ and covariance $\bm \Sigma$.

\renewcommand{\arraystretch}{2}

\begin{sidewaystable}
\caption{Gaussian distribution parameters for distributions of base layer parameters in SAR image model equations (\ref{sar-app-eq:samplestep-basic2}) and (\ref{sar-app-eq:samplestep-basic3})}
\label{sar-table:gaussian-parameters-inference}
\centering
\begin{tabular}{|c|c|c|c|c|c|}
\hline
{Component} & {Variable} & {Mean} & {Covariance} & {Parameters} \\
\hline\hline
\multirow{2}{*}{Background} & \multirow{2}{*}{$\sarBv$} & \multirow{2}{*}{${\bm \Lambda}\sarRvf$} & \multirow{2}{*}{$(\eyeKK-{\bm \Lambda})\GammaB(p)$}  & $\sarRvf = \sum_{i=1}^N  (\sarIv ./ \sarHv - \sarXv - \sarDG\sarGv - \sarDM\sarMv)$\\
& & & & ${\bm \Lambda}=\GammaB(p)\left(\GammaB(p)+\GammaV\right)^{-1}$\\
\hline
\multirow{3}{*}{Speckle, Target} & \multirow{3}{*}{$\left[\begin{array}{c}\sarXv\\\sarMv \end{array}\right]$} & \multirow{3}{*}{$\bm \Lambda\sarRv$} & \multirow{3}{*}{${\bm \Sigma}_{11}-{\bm \Lambda}{\bm \Sigma}_{12}$}  & ${\bm \Sigma}_{11} = \left[\begin{array}{cc} \GammaX(p) & 0 \\ 0 & \sarDM\GammaM\end{array}\right], {\bm \Sigma}_{12} = \left[\begin{array}{c} \GammaX(p) \\ \sarDM\GammaM\end{array}\right] $\\
& & & & $\sarRv = \sarIv ./ \sarHv - \sarBv - \sarDG\sarGv$\\
& & & & ${\bm \Lambda}={\bm \Sigma}_{12}^T\left(\GammaX(p)+\sarDM\GammaM+\GammaV\right)^{-1}$\\
\hline
\multirow{2}{*}{Glints} & \multirow{2}{*}{$\sarGv$} & \multirow{2}{*}{${\bm \Lambda}\sarRv$} & \multirow{2}{*}{$(\eyeKK-{\bm \Lambda}) \GammaG$}  & $\sarRv = \sarIv ./ \sarHv - \sarXv - \sarDM\sarMv - \sarBv$\\
& & & & ${\bm \Lambda}=\sarDG\GammaG\left(\sarDG\GammaG+\GammaV\right)^{-1}$\\
\hline
\end{tabular}
\end{sidewaystable}
\renewcommand{\arraystretch}{1.0}

\renewcommand{\arraystretch}{1.5}
\begin{sidewaystable}
\caption{Bernoulli distribution parameters for distributions of indicator variables in equations (\ref{sar-app-eq:samplestep-basic2}) and (\ref{sar-app-eq:samplestep-basic3})}
\label{sar-table:bernoulli-parameters-inference}
\centering
\begin{tabular}{|l|c|c|c|c|}
\hline
Component & Variable & {$f(\delta =1|-)$} & {$z_d$} & {$\underline{r}^{(p)}$}\\
\hline\hline
Glints & $\sarDG$ & $\left[1 + \frac{1-\pi_{f}^G(p)}{\pi_{f}^G(p)}\frac{z_0}{z_1}   \right]^{-1}$ & $\normpdf(\underline{r}^{(p)}/N;{\bm 0},d\GammaG+\GammaV)$ & $\sum_{i=1}^N \left(\sarIv ./ \sarHv - \sarBv - \sarXv - \sarDM\sarMv\right)$\\
\hline
Targets & $\sarDM$ & $\left[1 + \frac{1-\pi_{f,i}^M(p)}{\pi_{f,i}^M(p)}\frac{z_0}{z_1}   \right]^{-1}$ & $\normpdf(\underline{r}^{(p)};
{\bm 0},d\GammaM+\GammaV)$ & $\sarIv ./ \sarHv - \sarBv - \sarXv - \sarDG\sarGv$\\
\hline
\end{tabular}
\end{sidewaystable}
\renewcommand{\arraystretch}{1.0}

\subsection{Calibration coefficients}
For this thesis, we assumed that pixels within a subset $Z_g \subset \set{1,2,\dots,P}$ share the same calibration constant so that 
\begin{equation}
h_{k,f,i}^{(p)} = z_{k,f,i}(g),\qquad \forall p\in Z_g.
\end{equation}
with $z_{k,f,i}(g) \sim \cn{1}{(\sigma^H)^2}$.  In our formulation (and dropping the $(k,f,i)$ indices for simplicity) we have measurements of the form
\begin{equation}
i^{(p)} = z(g) (l^{(p)} + s^{(p)} + v^{(p)}),\qquad \forall p\in Z_g.
\end{equation}
Define $y^{(p)} = l^{(p)} + s^{(p)}$ which is a known quantity in our Gibbs sampling inference step.  Moreover, we assume that $|y^{(p)}|\gg |v^{(p)}|$ so that for any $p\in Z_g$ we have
\begin{equation}
\begin{split}
i^{(p)} &= z(g) \left( y^{(p)} + v^{(p)} \right)\\
&\approx  z(g) y^{(p)} + E[z(g)] v^{(p)} \\
&= z(g) y^{(p)} + v^{(p)}
\end{split}
\end{equation}
Note that given $y^{(p)}$ (as in the Gibbs sampling step), we have the situation where $i^{(p)}$, $z(g)$, and $v^{(p)}$ are all Gaussian distributed random variables.  Thus, the conditional distribution of $z(g)$ is also Gaussian with:
\begin{align}
z(g)|{\bm y} &\sim \cn{\mu_{z}(g)}{\sigma_z^2(g)}\\
\mu_z(g) &= 1 + \left[\frac{(\sigma^H)^2}{(\sigma^V)^2+{\bm y}^H{\bm y}(\sigma^H)^2}\right]{\bm y}^H\left({\bm i}-{\bm y}\right)\\
\sigma_z^2(g) &= \frac{(\sigma^H)^2(\sigma^V)^2}{(\sigma^V)^2 + {\bm y}^H{\bm y}(\sigma^H)^2}
\end{align}
where ${\bm y} = \set{y^{(p)}}_{p\in Z_g}$ and ${\bm i} = \set{i^{(p)}}_{p\in Z_g}$.  Note that when $(\sigma^H)^2$ is large, then maximum likelihood inference in this case yields the least-squares solution for $z(g)$.

\subsection{Object class assignment}
In this model, we assume that each pixel can be assigned to one of $J$ possible classes.   We assume that the number $J$ is known a priori and do not consider the details involved in the merging or splitting of object classes here.  More detailed models (such as the so-called Indian Buffet processes) can also estimate the number of classes directly from the data.

In this model, inference on class assignment is straightforward given the distributions (i.e., covariance matrices) for each class.  Define the matrices
\begin{equation}
{\bm b}^{(p)} = \left[
\begin{array}{c}
\underline{b}_1^{(p)}\\\underline{b}_2^{(p)}\\\vdots\\\underline{b}_F^{(p)}\end{array}\right]\in \mathbb{C}^{F\times K}
\qquad
{\bm x}^{(p)} = \left[
\begin{array}{c}
\underline{x}_{1,1}^{(p)}\\\underline{x}_{1,2}^{(p)}\\\vdots\\\underline{x}_{F,N}^{(p)}\end{array}\right]\in \mathbb{C}^{FN\times K}
\end{equation}
Then the probability that pixel $p$ belongs to class $j$ is given by:
\begin{equation}
{w}_j^{(p)} \trieq \Pr(\mathrm{pixel}\ p\mathrm{\ has\ class\ }j) = \exp\{T_B + T_X + q_j\}
\end{equation}
where $q_j$ is the prior probability of class $j$ and 
\begin{align}
T_B &= -\mathrm{trace}\left([\GammaB(j)]^{-1}{\bm b}^{(p)}({\bm b}^{(p)})^H\right) - \frac{F}{2} \log|\GammaB(j)| - KF \log(2\pi)\\
T_X &= -\mathrm{trace}\left([\GammaX(j)]^{-1}{\bm x}^{(p)}({\bm x}^{(p)})^H\right) - \frac{FN}{2} \log|\GammaX(j)| - KFN \log(2\pi)
\end{align}
Then the class assignment to pixel $p$ is the single location in $\underline{c}^{(p)}$ with value equal to one, where
\begin{equation}
\underline{c}^{(p)} \sim \mathrm{Multinomial}(1; \underline{w}^{(p)})
\end{equation}
Note that we can improve upon this model by allowing the probabilities for pixel $p$ to vary spatially (i.e., pixels are likely to share the same class with neighboring pixels).  One simple way to include this information is to let
\begin{equation}
\underline{m}^{(p)} = \underline{w}^{(p)} \ast \underline{g}_{HMM}^{(p)}
\end{equation}
where $ \underline{g}_{HMM}^{(p)}$ is some filter for averaging nearby pixels and $\ast$ is the convolution operator (assumed to be supported on the same set of pixels as $\underline{w}^{(p)}$).  Then we draw 
\begin{equation}
\underline{c}^{(p)} \sim \mathrm{Multinomial}(1; \underline{m}^{(p)})
\end{equation}
\subsection{Hyper-parameters}

In this model, we have three types of hyper-parameters that need to be estimated: covariance matrices (or variances), indicator probabilities, and object class probabilities.  In all cases, the distribution of these parameters depend on test statistics of much smaller dimension that $P$.

\subsubsection{Covariance matrix inference}
We model the covariance matrices for the Normal distributions in two ways: (1) for the stationary components (background, speckle, and glints), we model the covariance matrix as a random variable; and (2) for the other components (targets, additive noise, calibration coefficients), we assume independence among the antennas.  In particular, consider a random vector of $K$ elements, $\underline{w}$, with
\begin{align}
\underline{w} &\sim {\mathcal{CN}}\left({\bm 0},{\sigma^2{\bm \Gamma}_\rho}\right)\\
\sigma^2 &\sim \mathrm{InvGamma}(a_\sigma,b_\sigma)
\end{align}
Then in the stationary case, we have
\begin{align}
{\bm \Gamma}_\rho &\sim \mathrm{InvWishart}\left(a_{\Gamma}((1-\rho)\eyeKK + \rho\oneK\oneK^T), \nu_\Gamma\right)\\
\rho &\sim \mathrm{Beta}(a_\rho,b_\rho)
\end{align}
and in the independent case, we have
\begin{align}
{\bm \Gamma}_\rho = \eyeKK
\end{align}
First consider the case where ${\bm \Gamma}_\rho$ is a random variable.  Assume that we have $n$ independent samples of $\underline{w}$, which we refer to as ${\bm W} =\mathrm{vec}\set{\underline{w}}$.  Then, we consider a Gibbs sampling procedure:
\begin{align}
\mathrm{Sample}&\sim f(\sigma^2|{\bm W},{\bm \Gamma}_\rho,\rho)\\
\label{sar-app-eq:sample-step-covariance}\mathrm{Sample}&\sim f({\bm \Gamma}_\rho,\rho|{\bm W},\sigma^2)
\end{align}
Let $\tau=1/\sigma^2 \sim \mathrm{Gamma}(a_\sigma,b_\sigma)$. Then
\begin{equation}
\begin{split}
f(\tau|{\bm W},{\bm \Gamma}_\rho,\rho)&\propto f({\bm W}|\tau,{\bm \Gamma}_\rho) f(\tau)\\
&\propto \left[\tau^{n/2}\exp\set{-\frac{\tau}{2}\mathrm{trace}({\bm \Gamma}_\rho^{-1}{\bm W}{\bm W}^H)}\right] \left[\frac{(b_\sigma)^{a_\sigma}}{\Gamma(a_\sigma)}\tau^{a_\sigma-1}\exp\set{-b_\sigma\tau}\right]\\
&\propto \frac{(b')^{a'}}{\Gamma(a')}\tau^{a'-1}\exp\set{-b'\tau}
\end{split}
\end{equation}
where
\begin{equation}
\label{sar-app-eq:inverseGamma-posterior-general}
\begin{split}
a' &= a_\sigma + \frac{n}{2}\\
b' &= b_\sigma + \frac{\mathrm{trace}({\bm \Gamma}_\rho^{-1}{\bm W}{\bm W}^H)}{2}
\end{split}
\end{equation}
This demonstrates that in this situation, $\sigma^2$ has an Inverse-Gamma distribution with parameters $a'$ and $b'$.  Note that in the case where ${\bm \Gamma}_\rho=\eyeKK$, then the posterior parameters are given by
\begin{align}
a' &= a_\sigma + \frac{n}{2}\\
b' &= b_\sigma + \frac{\mathrm{trace}({\bm W}{\bm W}^H)}{2}
\end{align}
Thus, in the Gibbs sampling procedure, the variance parameter $\sigma^2$ is Inverse-Gamma distributed whether or not ${\bm \Gamma}_\rho$ is modeled as a random variable.  Table \ref{sar-table:variance-parameters-inference} provides the posterior Inverse Gamma distribution parameters for the variance parameters in our model, where $\mathrm{vec}\set{\cdot}$ refers to the vectorization operator.  

\renewcommand{\arraystretch}{1.5}
\begin{sidewaystable}
\caption{Inverse Gamma distribution parameters for distributions of variances and covariance matrix estimates}
\label{sar-table:variance-parameters-inference}
\centering
\begin{tabular}{|l|c|c|c|c|}
\hline
{Component} & {Variable} & {$a'$} & {$b'$} & {Other parameters}\\
\hline\hline
Background & $(\sigma^B(j))^2$ & $a_\sigma + \frac{M}{2} $ & $\frac{\mathrm{trace}({\bm \Gamma}_\rho^{-1}{\bm W}{\bm W}^H)}{2}$ & $M = F|Q_j| , {\bm W} = \mathrm{vec}\set{\sarBv}_{f,p\in Q_j}$\\
\hline
{Speckle} & $(\sigma^X(j))^2$ & $a_\sigma + \frac{M}{2} $ & $\frac{\mathrm{trace}({\bm \Gamma}_\rho^{-1}{\bm W}{\bm W}^H)}{2}$ & $M = NF|Q_j| , {\bm W} = \mathrm{vec}\set{\sarXv}_{f,i,p\in Q_j}$\\
\hline
{Glints}&$(\sigma^G)^2$ & $a_\sigma + \frac{M}{2} $ & $\frac{\mathrm{trace}({\bm \Gamma}_\rho^{-1}{\bm W}{\bm W}^H)}{2}$ & $M = \sum_{f,p}\sarDG , {\bm W} = \mathrm{vec}\set{\sarGv}_{\set{p,f,i:\sarDG=1}}$\\
\hline
{Target}&$(\sigma^M)^2$ & $a_\sigma + \frac{M}{2} $ & $\frac{\mathrm{trace}({\bm W}{\bm W}^H)}{2}$ & $M = \sum_{f,i,p}\sarDM , {\bm W} = \mathrm{vec}\set{\sarMv}_{\set{p,f,i:\sarDM=1}}$\\
\hline
{Calibration}&$(\sigma^H)^2$ & $a_\sigma + \frac{M}{2} $ & $\frac{\mathrm{trace}({\bm W}{\bm W}^H)}{2}$ & $M = KFNP/|Z_g|,\quad {\bm W} = \mathrm{vec}\set{z_{k,f,i}(g)}_{k,f,i,g}$\\
\hline
{Additive noise}& $(\sigma^V)^2$ & $a_\sigma + \frac{M}{2} $ & $\frac{\mathrm{trace}({\bm W}{\bm W}^H)}{2}$ & $M = KFNP,\quad {\bm W} = \mathrm{vec}\set{\frac{i_{k,f,i}^{(p)}}{h_{k,f,i}^{(p)}}-l_{k,f,i}^{(p)}-{s_{k,f,i}^{(p)}}}_{k,f,i,p}$\\
\hline
\multicolumn{5}{l}{\small{Note:  The set $Q_j$ is defined as  $Q_j\trieq\set{p:c^{(p)}=j}$}}
\end{tabular}
\end{sidewaystable}
\renewcommand{\arraystretch}{1.0}

For the background, speckle, and glint components, we also need to sample the coherence parameter $\rho$ and correlation matrix ${\bm \Gamma}_\rho$.  
Let $\tilde{\bm W}={\bm W}/\sigma$ be our observed measurements given $\sigma^2$ as given by equation (\ref{sar-app-eq:sample-step-covariance}).  Define $\tilde{\bm \mu}={\bm \mu}/\sigma$.  Then we have
\begin{equation}
\begin{split}
{\tilde{\bm W} | \left({\bm \Gamma}_\rho, \sigma^2, \rho\right)} &\sim \cn{\tilde{\bm \mu}}{{\bm \Gamma}_\rho}\\
{\bm \Gamma}_\rho &\sim \mathrm{InvWishart}\left( [\rho\oneK\oneK^T+(1-\rho)\eyeKK](\nu - K - 1), \nu\right)\\
\rho &\sim \mathrm{Beta}(a_\rho,b_\rho)
\end{split}
\end{equation}
Note that this is in the form of the Multivariate-Normal-Inverse-Wishart conjugate distribution given $\rho$.  This leads to the well known posterior parameters:
\begin{equation}
\label{sar-app-eq:Gamma-posterior-step}
{\bm \Gamma}_\rho | \left(\tilde{\bm W}, \sigma^2, \rho\right) \sim \mathrm{InvWishart}({\bm \Lambda}_\rho a_\Gamma+\sum\limits_{m=1}^n \left(\underline{y}_m-\tilde{\bm \mu}\right)\left(\underline{y}_m-\tilde{\bm \mu}\right)^H,\nu_\Gamma + n)
\end{equation}
where ${\bm \Lambda}_\rho = \rho\oneK\oneK^T+(1-\rho)\eyeKK$.  Ideally, we would like to sample both ${\bm \Gamma}_\rho$ and $\rho$ jointly.  Even though we can simply sample from equation (\ref{sar-app-eq:Gamma-posterior-step}), the same is not true for the density
\begin{equation}
\rho | \left(\tilde{\bm W}, \sigma^2, \rho\right)
\end{equation}
which is required in order to jointly sample these parameters.  Fortunately, we know that
\begin{equation}
f({\bm \Gamma}_\rho,\rho | \tilde{\bm W}, \sigma^2) \propto f({\bm \Gamma}_\rho | \tilde{\bm W}, \sigma^2,\rho) f(\rho)
\end{equation}
which is easily evaluated since we have closed form functions for both of these densities.  Thus, we can use Metropolis-Hastings to sample $\rho$ and ${\bm \Gamma}_\rho$.

\subsubsection{Indicator probabilities}
In the basic model where the indicator Beta distribution parameters do not depend spatially or temporally, then the posterior indicator probabilities for
\begin{align}
\delta &\sim \mathrm{Bernoulli}(\pi)\\
\pi &\sim \mathrm{Beta}(a_\pi,b_\pi)
\end{align}
are given by
\begin{align}
\pi | \delta \sim \mathrm{Beta}(a_\pi + \delta, b_\pi + (1-\delta))
\end{align}
Note that we can modify $a_\pi$ and $b_\pi$ as in Section  \ref{sar-subsection:indicator-prob}.  However, the posterior inference for the probabilities is identical by replacing $a_\pi$ and $b_\pi$ by their spatiotemporally varying version.

\subsubsection{Object class probabilities}
We use a Multinomial-Dirichlet conjugate pair to determine object class assignments, where the class probabilities $\underline{q}$ have a prior Dirichlet distribution with $c_j = 1/J$ for $j = 1,2,\dots,J$.  Then, after observing the class assignments, we can calculate the number of pixels in any class
\begin{equation}
N_j = |Q_j| = \left|\set{p: c^{(p)}=j}\right|
\end{equation}
Then the posterior distribution for the class probabilities is given by
\begin{equation}
\underline{q} | \set{N_j}_{j=1}^J \sim \mathrm{Gamma}\left(\underline{q} + \underline{N},1\right)/J
\end{equation}
where $[\underline{N}]_j = N_j$.

}


\bibliographystyle{IEEEtran}
\bibliography{References}
\begin{IEEEbiography}[{\includegraphics[width=1in,height=1.25in,clip,keepaspectratio]{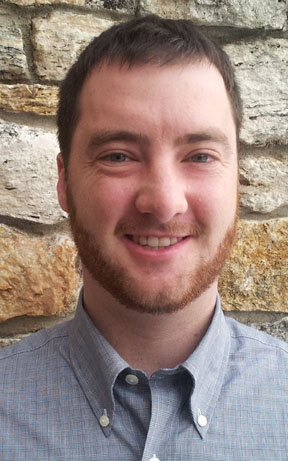}}]{Gregory Newstadt}
received the B.S. degrees (summa cum laude) from Miami University, Oxford, OH, in 2007 in Electrical Engineering and in Engineering Physics.  He also received the M.S.E in Electrical Engineering: Systems (2009), M.A. in Statistics (2012) and Ph.D. in Electrical Engineering: Systems (2013) degrees from the University of Michigan, Ann Arbor, MI.

He is currently a postdoctoral researcher and lecturer at the University of Michigan, Ann Arbor, MI, in Electrical Engineering (Systems).  His research interests include detection, estimation theory, target tracking, sensor fusion, and statistical signal processing.
\end{IEEEbiography}

\begin{IEEEbiography}[{\includegraphics[width=1in,height=1.25in,clip,keepaspectratio]{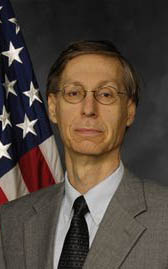}}]{Edmund Zelnio}
graduated from Bradley University, Peoria, Illinois, in 1975 and has pursued doctoral studies at The Ohio State University in electromagnetics and at Wright State University in signal processing.  He has had a 37 year career with the Air Force Research Laboratory (AFRL), Wright Patterson AFB, Ohio where he has spent 35 years working in the area of automated exploitation of imaging sensors primarily addressing synthetic aperture radar.  He is a former division chief and technical advisor of the Automatic Target Recognition Division of the Sensors Directorate in AFRL and serves in an advisory capacity to the Department of Defense and the intelligence community.  He is currently the director of the Automatic Target Recognition Center in AFRL.  He is the recipient of the 53rd DoD Distinguished Civilian Service Award and is a fellow of the Air Force Research Laboratory.
\end{IEEEbiography}

\begin{IEEEbiography}[{\includegraphics[width=1in,height=1.25in,clip,keepaspectratio]{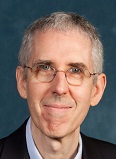}}]{Alfred O. Hero, III}
received the B.S. (summa cum laude) from Boston University (1980) and the Ph.D from Princeton University (1984), both in Electrical Engineering. Since 1984 he has been with the University of Michigan, Ann Arbor, where he is the R. Jamison and Betty Professor of Engineering. His primary appointment is in the Department of Electrical Engineering and Computer Science and he also has appointments, by courtesy, in the Department of Biomedical Engineering and the Department of Statistics. In 2008 he was awarded the the Digiteo Chaire d'Excellence, sponsored by Digiteo Research Park in Paris, located at the Ecole Superieure d'Electricite, Gif-sur-Yvette, France. He has held other visiting positions at LIDS Massachussets Institute of Technology (2006), Boston University (2006), I3S University of Nice, Sophia-Antipolis, France (2001), Ecole Normale Sup\'erieure de Lyon (1999), Ecole Nationale Sup\'erieure des T\'el\'ecommunications, Paris (1999), Lucent Bell Laboratories (1999), Scientific Research Labs of the Ford Motor Company, Dearborn, Michigan (1993), Ecole Nationale Superieure des Techniques Avancees (ENSTA), Ecole Superieure d'Electricite, Paris (1990), and M.I.T. Lincoln Laboratory (1987 - 1989).

Alfred Hero is a Fellow of the Institute of Electrical and Electronics Engineers (IEEE). He has been plenary and keynote speaker at major workshops and conferences. He has received several best paper awards including: a IEEE Signal Processing Society Best Paper Award (1998), the Best Original Paper Award from the Journal of Flow Cytometry (2008), and the Best Magazine Paper Award from the IEEE Signal Processing Society (2010). He received a IEEE Signal Processing Society Meritorious Service Award (1998), a IEEE Third Millenium Medal (2000) and a IEEE Signal Processing Society Distinguished Lecturership (2002). He was President of the IEEE Signal Processing Society (2006-2007). He sits on the Board of Directors of IEEE (2009-2011) where he is Director Division IX (Signals and Applications).

Alfred Hero's recent research interests have been in detection, classification, pattern analysis, and adaptive sampling for spatio-temporal data. Of particular interest are applications to network security, multi-modal sensing and tracking, biomedical imaging, and genomic signal processing.
\end{IEEEbiography}

\end{document}